\documentclass[fleqn,usenatbib]{mnras}

\usepackage{newtxtext,newtxmath}

\usepackage[T1]{fontenc}
\usepackage{ae,aecompl}

\usepackage{mathtools}
\usepackage{graphicx}	
\usepackage{float}
\usepackage{amsmath}	
\usepackage{amssymb}	

\graphicspath{{./}{figures/}}

\title[Testing HOD]{Testing the Accuracy of Halo Occupation Distribution Modelling using Hydrodynamic Simulations}

\author[Beltz-Mohrmann et al.]{
Gillian D. Beltz-Mohrmann$^{1}$\thanks{E-mail: gillian.d.beltz-mohrmann@vanderbilt.edu (GDBM)},
Andreas A. Berlind$^{1}$, Adam O. Szewciw$^{1}$
\\
$^{1}$ Department of Physics and Astronomy, Vanderbilt University, Nashville, TN 37235, USA}

\date{Accepted XXX. Received YYY; in original form ZZZ}

\pubyear{2019}

\begin{document}
\label{firstpage}
\pagerange{\pageref{firstpage}--\pageref{lastpage}}
\maketitle

\begin{abstract}
Halo models provide a simple and computationally inexpensive way to investigate the connection between galaxies and their dark matter haloes. However, these models rely on the assumption that the role of baryons can be easily parametrized in the modelling procedure. We aim to examine the ability of halo occupation distribution (HOD) modelling to reproduce the galaxy clustering found in two different hydrodynamic simulations, Illustris and EAGLE. For each simulation, we measure several galaxy clustering statistics on two different luminosity threshold samples. We then apply a simple five parameter HOD, which was fit to each simulation separately, to the corresponding dark matter only simulations, and measure the same clustering statistics. We find that the halo mass function is shifted to lower masses in the hydrodynamic simulations, resulting in a galaxy number density that is too high when an HOD is applied to the dark matter only simulation. However, the exact way in which baryons alter the mass function is remarkably different in the two simulations. After applying a correction to the halo mass function in each simulation, the HOD is able to accurately reproduce all clustering statistics for the high luminosity sample of galaxies. For the low luminosity sample, we find evidence that in addition to correcting the halo mass function, including spatial, velocity, and assembly bias parameters in the HOD is necessary to accurately reproduce clustering statistics.
\end{abstract}

\begin{keywords}
cosmology: large-scale structure of Universe --- cosmology: dark matter --- galaxies: haloes --- galaxies: clusters: general --- galaxies: statistics
\end{keywords}



\section{Introduction}
Studying the connection between galaxies and the dark matter haloes in which they reside is one of the keys to understanding galaxy formation and evolution, as well as constraining cosmological models. In recent years, using hydrodynamic simulations has become a popular method for investigating this connection \citep[e.g.][]{2014Natur.509..177V}. However, these simulations are computationally expensive, and are thus ill-suited for exploring a large parameter space. Moreover, different hydrodynamic simulations produce different results; we currently lack a consensus on the correct gas physics prescriptions to use. 

By contrast, dark matter only simulations are much less computationally expensive, and although the only physics involved is gravity, they still allow us to predict the large-scale distribution of dark matter as well as the statistical properties of dark matter haloes in the universe. One can then adopt an empirical rather than an {\it ab-initio} approach and employ a halo model in order to connect galaxies to the dark matter distribution. Halo models are a broad class of models based on the assumption that galaxies form and live inside dark matter haloes. With a few free parameters that can be fit to clustering observations, one can connect galaxies to haloes, thus quantitatively modelling galaxy clustering on small scales while bypassing the need for a complete understanding of galaxy formation physics.

The earliest halo models to describe galaxy clustering were the analytic models of \citet{1952ApJ...116..144N}, \citet{1974A&A....32..197P}, and \citet{1977ApJ...217..331M}. Later, \citet{1997MNRAS.286..795K,1999MNRAS.303..188K}, and \citet{1999MNRAS.305L..21B} showed that semi-analytic models could be used to predict galaxy clustering by combining the results from N-body simulations with theories for the formation and evolution of galaxies within haloes. Soon thereafter, \citet{1998ApJ...494....1J} and \citet{2000MNRAS.311..793B} found that galaxy clustering merely depends on halo occupation statistics as a function of halo mass, potentially sidestepping the need to model galaxy formation altogether. Subsequently, several papers \citep[e.g.][]{2000ApJ...543..503M,2000MNRAS.318.1144P,2000MNRAS.318..203S,2001ApJ...546...20S,2001MNRAS.325.1288S,2001ApJ...550L.129W,2002PhR...372....1C} expanded on the work of \citet{1991ApJ...381..349S} to combine both halo properties and occupation statistics to successfully predict the galaxy correlation function and power spectrum.

A key ingredient of the halo model is the Halo Occupation Distribution (HOD), which defines the bias of a population of galaxies by the conditional probability that a dark matter halo of virial mass M contains N galaxies, together with prescriptions that specify the relative spatial and velocity distributions of galaxies and dark matter within haloes \citep{2002ApJ...575..587B,2003ApJ...593....1B}. These relations can be parametrized with various degrees of freedom. However, most studies have used simple formulations of the HOD, with at most five free parameters that specify the mean occupation number of galaxies, along with the assumptions that galaxies trace dark matter inside haloes. This type of HOD model, as proposed by \citet{2005ApJ...633..791Z}, has become the `standard' in halo modelling studies.

Halo models have been used to model galaxy clustering in many galaxy redshift surveys, including the Sloan Digital Sky Survey \citep[SDSS;][]{2000AJ....120.1579Y}, the 2dF Galaxy Redshift Survey \citep[2dFGRS;][]{2001MNRAS.328.1039C}, the 6dF Galaxy Redshift Survey \citep[6dfGRS;][]{2004MNRAS.355..747J}, and the SDSS III Baryon Oscillation Spectroscopic Survey \citep[BOSS;][]{2013AJ....145...10D}. Many studies have used halo models to investigate the two-point correlation function of both low redshift galaxies \citep[e.g.][]{2003MNRAS.346..186M, 2004ApJ...608...16Z, 2005MNRAS.361..415C, 2005ApJ...631...41T, 2005ApJ...630....1Z, 2011ApJ...736...59Z, 2012ApJ...749...83W, 2013MNRAS.429.3604B, 2015ApJ...806..125P} as well as high redshift galaxies \citep[e.g.][]{2002MNRAS.329..246B, 2002ApJ...577....1M, 2004MNRAS.347..813H, 2004ApJ...610...61Z, 2006ApJ...642...63L, 2010ApJ...709...67T, 2013MNRAS.429.2333J, 2014MNRAS.438..825K} (as cited in \citet{2018MNRAS.478.1042S}). 

Some previous works \citep[e.g.][]{2011ApJ...736...59Z} have found statistical tension between predictions of the halo model and the real universe when fitting to galaxy clustering measurements in the SDSS. However, these works rely on analytic halo models that do not adequately control for systematic errors in the modelling procedure, making it difficult to interpret the goodness of fit results. Recently, \citet{2018MNRAS.478.1042S} used a ``fully numerical mock-based methodology" to test the standard $\Lambda$CDM + halo model against the clustering of SDSS DR7 galaxies. Their procedure carefully controlled for systematic errors, allowing them to interpret the goodness of fit of their model. They measured the projected correlation function, group multiplicity function, and galaxy number density, and found that while the model could successfully fit each statistic separately, it was unable to fit them simultaneously. Their best-fitting model was able to reproduce the clustering of low luminosity galaxies, but revealed a 2.3$\sigma$ tension with the clustering of high luminosity galaxies, indicating a possible problem with the `standard' HOD model.

There are several assumptions built into the standard HOD model that could be incorrect. First, the HOD framework relies on the assumption that cosmology and gravity alone govern the dark matter halo distribution. However, it has been shown that gas physics can also affect the properties of haloes \citep[e.g.][]{2012MNRAS.423.2279C,2016MNRAS.456.2361B}. Second, the HOD typically assumes that the occupation of galaxies is solely based on halo mass, and does not depend on secondary halo properties like halo concentration or age. This ignores the possibility that galaxy clustering may be affected by the phenomenon known as assembly bias \citep{2005MNRAS.363L..66G,2006ApJ...652...71W,2007MNRAS.374.1303C,2018arXiv180906424P,2018MNRAS.475.4411S,2018arXiv181211210X,2018ApJ...853...84Z,2019MNRAS.484.1133C}. Finally, most HOD modelling assumes that galaxy positions and velocities within haloes trace the underlying distribution of dark matter.

\citet{2014MNRAS.443.3044Z} examined the extent to which the presence of assembly bias could lead to systematic errors in halo occupation statistics inferred from galaxy clustering. The authors constructed two sets of realistic mock galaxy catalogues with identical HODs: one with assembly bias and one with assembly bias removed. They then fit standard HODs to the galaxy clustering in each catalogue, and found that in the case where assembly bias was removed, the inferred HODs agreed with the true HODs, but when assembly bias was included, the inferred HODs showed significant systematic errors.

\citet{2016MNRAS.460.2552H} introduced a new class of HOD models, known as `decorated HODs', designed to incorporate parameters for assembly bias in halo occupation distribution models. The authors used these new models to characterize the impact of assembly bias on clustering statistics, and found that for SDSS-like samples, assembly bias can affect galaxy clustering by up to a factor of 2 on 200 kpc scales. They also found that on small scales (r < 1 Mpc) assembly bias generally enhances clustering, but on large scales it can either increase or decrease clustering. \citet{2019ApJ...872..115V} and \citet{2019MNRAS.485.1196Z} applied this decorated HOD model to galaxies in the SDSS DR7 and found evidence of galaxy assembly bias for some luminosity samples.

Regarding the spatial distribution of galaxies within haloes, the HOD often uses random dark matter particles to assign positions and velocities to galaxies, or otherwise assumes a dark matter density profile for galaxies \citep[e.g.][NFW]{1997ApJ...490..493N}. This does not account for the possibility that galaxies might not move like dark matter due to phenomena such as mergers, tidal stripping, and dynamical friction, leading to effects like spatial and velocity bias. Both \citet{2012ApJ...749...83W} and \citet{2015ApJ...806..125P} used halo models to predict the very small-scale clustering of galaxies in the SDSS, and found that more luminous galaxies do not trace underlying dark matter distributions of their haloes, indicating the presence of spatial bias. \citet{2015MNRAS.446..578G} looked at galaxy clustering in SDSS DR11 and found observational evidence for central velocity bias (i.e. that central galaxies on average are not at rest with respect to their host haloes) as well as satellite velocity bias (i.e. in this case, that luminous satellite galaxies move more slowly than the dark matter). In a subsequent paper, \citet{2015MNRAS.453.4368G} modelled the projected and redshift-space two-point correlation functions of galaxies in SDSS DR7, and similarly found that luminous central galaxies and faint satellite galaxies exhibit velocity bias. Furthermore, they found that their measurements could be successfully interpreted within an extended HOD framework that includes central and satellite velocity bias parameters to describe the motions of galaxies within haloes.

\begin{table*}
	\centering
	\caption{Simulation parameters. The columns show (from left to right): simulation name, box size, number of dark matter particles, dark matter particle mass (for the hydrodynamical run), redshift used, and cosmological parameters. The dark matter particle mass for Illustris-2-Dark is $4.2 \times 10^7 (h^{-1} M_\odot)$, and for EAGLE Dark it is $7.5 \times 10^6 (h^{-1} M_\odot)$.}
	\label{tab:sim_table}
	\begin{tabular}{ccccccccccc}
		\hline
		Simulation & $L_\mathrm{box} (h^{-1} \mathrm{Mpc})$ & $N_\mathrm{DM}$ & $m_\mathrm{DM} (h^{-1} M_\odot)$ & $z$ & $h$ & $\Omega_{m}$ & $\Omega_{\Lambda}$ & $\Omega_b$ & $\sigma_8$ & $n_s$\\
		\hline
		Illustris-2 & 75 & $910^3$ & $3.5 \times 10^7$ & 0.13 & 0.704 & 0.2726 & 0.7274 & 0.0456 & 0.809 & 0.963\\
		EAGLE & 67.77 & $1504^3$ & $6.6 \times 10^6$ & 0.101 & 0.6777 & 0.307 & 0.693 & 0.04825 & 0.8288 & 0.9611\\
		\hline
	\end{tabular}
\end{table*}

\citet{2014MNRAS.442.1930P} investigated how well an HOD model could reproduce the two-point clustering of galaxies in several semi-analytic models, and found that the HOD failed to reconstruct the galaxy bias for low mass haloes, indicating the presence of assembly bias. They also found that clustering shows some dependence on the substructure of the host halo. Subsequently, \citet{2017A&A...598A.103P} further compared the HOD model to semi-analytic models, and found that using local density rather than halo mass in the HOD model was a better predictor of galaxy bias.

In this paper we use hydrodynamic simulations of galaxy formation to investigate the extent to which all these built-in assumptions to the standard HOD model can affect galaxy clustering statistics. Although previous works \citep[e.g.][]{2018MNRAS.480.3978A,2019arXiv190508799B} have used hydrodynamic simulations to investigate variations in halo occupancy with environment, concentration, and formation time, none have looked at the impact of the assumptions of the HOD on galaxy clustering statistics compared to clustering in hydrodynamic simulations. Additionally, previous works have not looked at a wide variety of clustering statistics, nor have they compared bias effects across multiple different hydrodynamic simulations. 

In this work, we focus on two different hydrodynamic simulations, as well as two different luminosity threshold samples of galaxies. We measure several different galaxy clustering statistics on each of our samples. We then fit a five parameter HOD model to each simulation and sample, and apply these models to the corresponding dark matter only simulations. We then measure the same galaxy clustering statistics on our HOD galaxies as we did on our hydrodynamic galaxies. We examine the accuracy with which we can predict galaxy clustering using our HOD modelling framework, as compared to the full hydrodynamic simulations. Finally, we investigate how we might expand the HOD model to include effects like assembly, spatial, and velocity bias in order to increase the accuracy of the model. We note that our analysis strictly compares HOD modeling to hydrodynamic simulations and not to real galaxy surveys. Therefore, conclusions should not be drawn about the accuracy of the clustering produced either by the simulations or the HOD models as compared to real observations. However, the conclusions that we draw about the need to add freedom to HOD models are still valid.

We discuss our simulations in Section~\ref{sims}, and our halo model in Section~\ref{hod}. In Section~\ref{stats} we discuss our clustering statistics, and in Section~\ref{fit} we discuss the accuracy of our model. In Section~\ref{hmf} we discuss our halo populations, and in Section~\ref{decorated} we discuss possible extensions to our HOD model. Finally, in Section~\ref{summary} we summarize our results and conclusions.

\section{Simulations} \label{sims}
We use two cosmological N-body simulations for our analysis: Illustris \citep{2015A&C....13...12N,2014MNRAS.444.1518V,2014Natur.509..177V,2014MNRAS.445..175G} and EAGLE \citep{2015MNRAS.446..521S,2016A&C....15...72M,2017arXiv170609899T,2005MNRAS.364.1105S,2015MNRAS.450.1937C}. The Illustris-2 simulation has a volume of $75^3 (h^{-3} \mathrm{Mpc}^3)$ and a dark matter particle mass of $3.5 \times 10^7 (h^{-1} M_\odot)$. The EAGLE simulation (RefL100N1504) has a volume of $67.77^3 (h^{-3} \mathrm{Mpc}^3)$ and a dark matter particle mass of $6.6 \times 10^6 (h^{-1} M_\odot)$. A summary of the simulation parameters can be found in Table~\ref{tab:sim_table}.

Each of these hydrodynamic simulations has a corresponding dark matter only (DMO) counterpart, derived from the same cosmology and initial conditions. These two simulations are ideal for our analysis because they have high enough resolutions for the galaxies we are interested in, as well as large enough volumes to accurately measure clustering statistics out to $10 h^{-1} \mathrm{Mpc}$ scales. We specifically choose to use Illustris-2 because the resolution of Illustris-3 is not quite high enough for our purposes, but the resolution of Illustris-1 is not necessary for the halo mass range that we are interested in. This is because in this work, the smallest haloes that we will ever populate with galaxies using our HOD model are on the order of $10^{11} (h^{-1} M_\odot)$. In Illustris-2-Dark, a halo of this size has about 2400 particles, so it is well-resolved. Additionally, such a small halo will only ever be assigned a central galaxy (if it is assigned a galaxy at all), and thus the only halo properties that we need to know are the position and velocity of the halo, which should be well-established with 2400 particles.

The Illustris simulation was performed with the moving-mesh code \texttt{AREPO}, while the EAGLE simulation was performed with the \texttt{GADGET-3} tree-SPH code, a modified version of the public \texttt{GADGET-2} simulation code. Both simulations employ models for star formation, stellar evolution, gas cooling and heating, supernovae feedback, black hole formation, and AGN feedback. According to \citet{2012MNRAS.423.1726S}, while \texttt{GADGET-3} and \texttt{AREPO} share the same sub-grid physics, their different numerical hydrodynamical techniques can lead to large discrepancies in their galaxies. In their tests, \texttt{GADGET-3} formed only about half as many stars as \texttt{AREPO}, and \texttt{AREPO} has a much higher gas and stellar mass fraction than \texttt{GADGET-3}. The benefit of using two simulations with different physics for our analysis is that we can compare our results from the two different simulations, providing us with some theoretical uncertainty on our results.

\begin{table*}
	\centering
	\caption{HOD parameters for each sample. The columns show (from left to right): the simulation name, the absolute magnitude limit for the SDSS sample whose number density we are matching, the absolute magnitude limit used in the case of the given simulation, the galaxy number density, the five best-fitting HOD parameters for that sample, and the corresponding reduced chi-square value.}
	\label{tab:hod_table}
	\begin{tabular}{cccccccccc}
		\hline
		Simulation & $M_r \mathrm{(SDSS)}$ & $M_r^\mathrm{lim}$ & $n_\mathrm{g} (h^{3}\mathrm{Mpc}^{-3})$ & $\mathrm{log} M_\mathrm{min}$ & $\sigma_{\mathrm{log}M}$ & $\mathrm{log} M_0$ & $\mathrm{log} M_1$ & $\alpha$ & $\chi^2/dof$\\
		\hline
		Illustris & -21 & -22.840 & 0.0012 & 12.681 & 0.532 & 12.296 & 13.635 & 0.994 & 0.908\\
		Illustris & -19 & -20.354 & 0.0149 & 11.500 & 0.180 & 11.659 & 12.590 & 0.979 & 8.560\\
		\hline
		EAGLE & -21 & -21.852 & 0.0012 & 12.767 & 0.504 & 12.467 & 13.799 & 1.000 & 1.498\\
		EAGLE & -19 & -19.695 & 0.0149 & 11.555 & 0.237 & 11.717 & 12.566 & 0.938 & 3.635\\
		\hline
	\end{tabular}
\end{table*}

We are interested in two different samples of galaxies: a ``high" luminosity sample, similar to that of the volume-limited SDSS DR7 \citep{2009ApJS..182..543A} $M_r<-21$ sample, and a ``low" luminosity sample, similar to that of the SDSS DR7 $M_r<-19$ sample. (We will refer to these samples as $M_r^{-21}$ and $M_r^{-19}$ henceforth.) We choose to use the $z=0.13$ snapshot of the Illustris simulation because it is the closest available redshift to the median redshift of the SDSS $M_r^{-21}$ sample ($z_\mathrm{med} = 0.132$). We choose the $z=0.101$ snapshot of the EAGLE simulation because it is also the closest available redshift to that of the SDSS DR7 $M_r^{-21}$ sample. The $M_r^{-19}$ luminosity threshold sample has a median redshift of $0.054$. For the EAGLE simulation, the closest available redshift is still the $z=0.101$ snapshot. Therefore, because the snapshot does not change for our analysis on the EAGLE simulation, we likewise chose not to change the snapshot for the Illustris simulation. However, there is little evolution between $z=0.13$ and $z=0.054$, and we do not compare our clustering statistics to those measured on SDSS data, so our choice of snapshot should not impact our results.

To create our galaxy samples, for each simulation we find the luminosity threshold that results in a galaxy number density equivalent to that of the SDSS datasets of interest (either $M_r^{-21}$ or $M_r^{-19}$). The luminosity threshold for each simulation and sample is given in Table~\ref{tab:hod_table}. We note that the luminosity thresholds are not exactly $-21$ or $-19$, which indicates that the luminosity functions in Illustris and EAGLE are not the same as that in the SDSS, nor are they the same as each other. (This discrepancy emphasizes the lack of consensus among hydrodynamic simulations, and thus the advantage of using HOD modeling with plenty of freedom to model galaxy clustering in the real Universe.) Thus, if we create our samples based on luminosity, our number density will be different than that of the SDSS samples. Therefore, we choose to use a different luminosity threshold to do an accurate number density comparison. We will still refer to the samples as the $M_r^{-21}$ and $M_r^{-19}$ samples. 

After setting the luminosity threshold, we then determine the number of remaining galaxies in each halo, and average in bins of halo mass. For the $M_r^{-21}$ samples we use 14 evenly spaced logarithmic bins between 11.9 and 14.52. For the $M_r^{-19}$ samples we use 20 evenly spaced logarithmic bins between 11.0 and 14.52. Our halo occupation distributions for each galaxy sample are shown in Figure~\ref{fig:bestfit_hod}. The Illustris samples are plotted in red, and the EAGLE samples are plotted in blue.

\section{Halo Occupation Modelling} \label{hod}
\subsection{The Halo Occupation Distribution}
\label{sec:hod}
The Halo Occupation Distribution framework governs the number, positions, and velocities of galaxies within a dark matter halo based on a few free parameters, which depend only on the mass of the halo. The version of the HOD that we utilize in this work is the five parameter `vanilla' HOD model of \citet{2007ApJ...667..760Z} (as cited in \citet{2018MNRAS.478.1042S}). Within their haloes, galaxies are split into centrals and satellites \citep{2004ApJ...609...35K,2005ApJ...633..791Z}. 

The mean number of central galaxies in a halo of mass $M$ is described by\footnote{Throughout this paper, $\mathrm{log}$ refers to $\mathrm{log}_{10}$.}
\begin{equation}
\langle N_\mathrm{cen} \rangle = \frac{1}{2}\bigg[1 + \mathrm{erf} \bigg(\frac{\mathrm{log} M - \mathrm{log}M_\mathrm{min}}{\sigma_{\mathrm{log} M}}\bigg)\bigg],
\end{equation}
where $M_\mathrm{min}$ is the mass at which half of halos host a central galaxy, $\sigma_{\mathrm{log} M}$ is the scatter around this halo mass, and $\mathrm{erf}(x)$ is the error function, $\mathrm{erf}(x)=\frac{2}{\sqrt{\pi}}\int_0^x \mathrm{exp}(-y^2)dy$. The central galaxy is always placed at the centre of the halo, and given the mean velocity of the halo (i.e. we assume that the central galaxy is at rest with respect to the halo).

We determine the number of satellite galaxies to place in each halo by drawing from a Poisson distribution with a mean given by
\begin{equation}
\langle N_\mathrm{sat} \rangle = \langle N_\mathrm{cen} \rangle \times \bigg(\frac{M - M_0}{M_1}\bigg)^\alpha,
\end{equation}
where $M_0$ is the halo mass below which there are no satellite galaxies, $M_1$ is the mass where haloes contain on average one satellite galaxy, and $\alpha$ is the slope of the power-law occupation function at high masses. Each satellite galaxy is assigned the position and velocity of a randomly chosen dark matter particle within the halo, i.e. we assume that satellite galaxies trace the spatial and velocity distribution of dark matter within the halo.

In summary, our HOD model contains five free parameters that control the number of galaxies in each halo as a function of halo mass. Our model assumes that all galaxies live inside dark matter haloes, and that the number of galaxies in a halo depends only on the mass of the halo and not on any other halo properties, such as age or concentration (i.e. there is no galaxy assembly bias). However, recent work \citep[e.g.][]{2014MNRAS.443.3044Z,2019ApJ...872..115V,2019MNRAS.485.1196Z} indicates that galaxy assembly bias is probably present in luminosity threshold samples, so this assumption is likely incorrect. 

Additionally, our model assumes that the number of satellite galaxies in each halo is governed by a Poisson distribution. However, results from simulations indicates that the scatter in the number of satellite galaxies at fixed halo mass is probably non-Poissonian \citep{2010MNRAS.406..896B,2015ApJ...810...21M}. In fact, \citet{2019arXiv190604298J} found that the HOD was best able to reproduce the spatial distribution of galaxies in a semi-analytical model when they used a negative binomial distribution to govern the number of satellite galaxies in a halo. 
\begin{figure*}
\centering
	\includegraphics[width=6in]{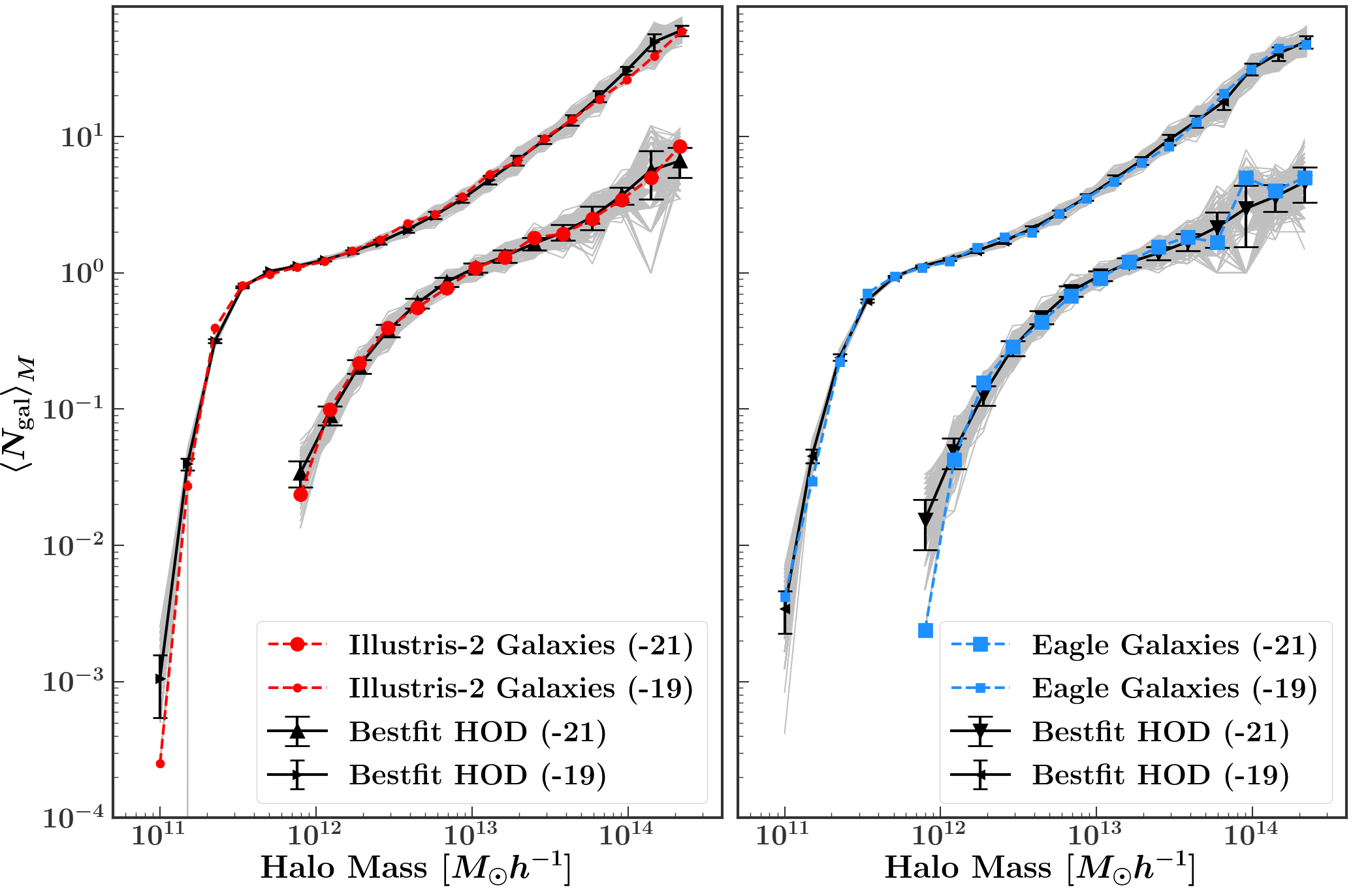}
    \caption{Best-fitting HOD for Illustris-2 (left) and EAGLE (right) galaxies. The Illustris-2 high luminosity ($M_r^{-21}$) galaxy sample is plotted with a solid red line, and the low luminosity ($M_r^{-19}$) sample is plotted with a dashed red line, while the EAGLE high luminosity sample is plotted with a solid blue line, and the low luminosity sample is plotted with a dashed blue line. The gray lines in each case show 300 realizations of the best-fitting HOD model for that sample. The black line and error bars represent the mean and standard deviation among these 300 realizations.}
    \label{fig:bestfit_hod}
\end{figure*}

Finally, our model assumes that the central galaxy in each halo lives at the centre of the halo and moves with the mean velocity of the halo (i.e. there is no central spatial or velocity bias), and that the satellite galaxies in each halo follow the spatial and velocity distribution of dark matter within the halo (i.e. there is no satellite spatial or velocity bias). However, observations suggest that both central and satellite galaxies probably do exhibit spatial bias \citep[e.g.][]{2012ApJ...749...83W,2015ApJ...806..125P} as well as velocity bias \citep[e.g.][]{2005MNRAS.361.1203V,2015MNRAS.446..578G, 2015MNRAS.453.4368G}. 

While we do use this standard `vanilla' HOD in our initial analysis, we will discuss variations and extensions of this model in Section~\ref{decorated}.

\subsection{Fitting the HOD}
\label{sec:bestfit}
Next, we need to determine the five parameters that best describe the HOD in each simulation and sample. We do this in the following way. We start with an initial guess for each parameter. Using this fiducial HOD model, we assign a number of central and satellite galaxies to the haloes in the hydrodynamic run of the simulation. (The halo mass that we use for this is the total friends-of-friends group mass, i.e. including dark matter as well as baryonic particles.) Because there is some random variation in the HOD modelling framework, we repeat this process 300 times in order to generate 300 different realizations of our fiducial HOD. We then determine the number of galaxies in each halo (averaged in bins of halo mass), in the same way that we did for the original galaxies in the simulation. We can then calculate a $\chi^2$ to assess how well our fiducial HOD model fits the simulation:
\begin{equation}
\chi^2 = \sum_i \frac{(D_i - M_i)^2}{\sigma_i^2},
\end{equation}
where $D_i$ is the number of galaxies in one halo mass bin from the simulation, $M_i$ is the number of galaxies in the same halo mass bin averaged over 300 realizations of our fiducial HOD model, and $\sigma_i$ is the standard deviation among the 300 different realizations of our fiducial HOD. We do this separately for centrals and satellites, and then sum over all of our halo mass bins. Based on this $\chi^2$, we adjust our fiducial HOD parameters and repeat this process. We use a Nelder-Mead optimization algorithm \citep{NeldMead65,neldermead,jones_scipy_2001} to minimize $\chi^2$.

In Table~\ref{tab:hod_table}, we list the luminosity thresholds for each sample, as well as the best-fitting HOD parameters for each simulation. Shown in Figure~\ref{fig:bestfit_hod} are the best fit HODs for each of our simulations and density samples. While the $M_r^{-21}$ samples in both simulations each achieved a $\chi^2/DOF$ of close to 1, the $M_r^{-19}$ samples are not fit as well by the HOD, particularly in Illustris. This could be an indication that the form of the HOD is not optimal for describing a low-luminosity galaxy sample, but it can easily describe a high-luminosity sample.

\begin{figure*}
\centering
	\includegraphics[width=6in]{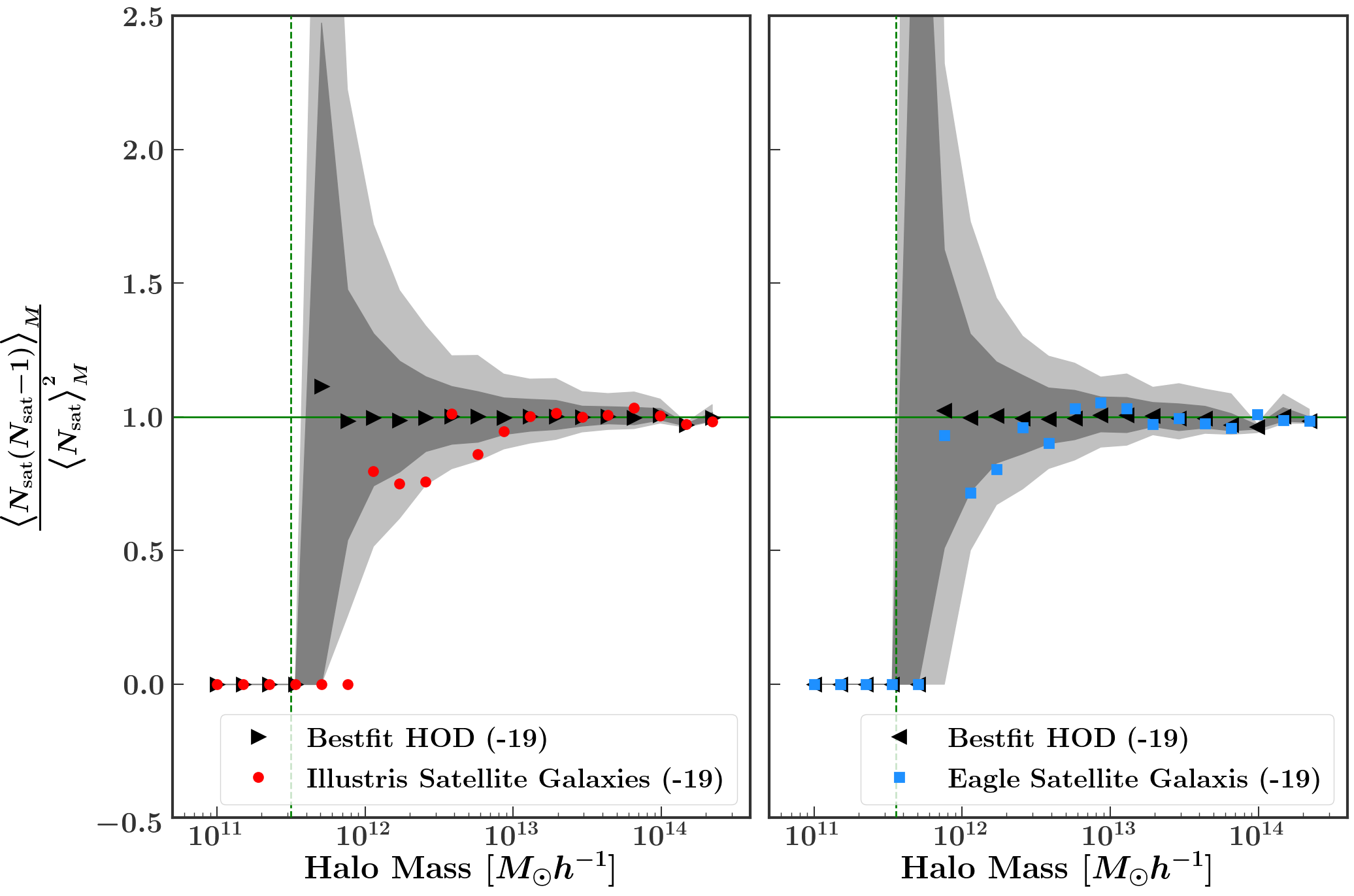}
    \caption{The second moment of the HOD for Illustris-2 $M_r^{-19}$ galaxies (red points, left) and EAGLE $M_r^{-19}$ galaxies (blue points, right). The dark and light gray shaded regions show the inner 68 and 95\% of the realizations of the best-fitting HOD model for that sample, and the black points are the median of the 300 realizations.}
    \label{fig:poisson}
\end{figure*}

One of the assumptions made in our modelling procedure is that the probability distribution governing the number of satellite galaxies in a halo is Poissonian. To investigate this assumption we examine the average number of satellite-satellite pairs per halo in bins of halo mass, $\big \langle N(N-1) \big \rangle_M$, or $\big \langle N^2 \big \rangle_M - \big \langle N \big \rangle_M$. A Poisson distribution of mean $\big \langle N \big \rangle$ has variance $\big \langle N^2 \big \rangle = \big \langle N \big \rangle^2 + \big \langle N \big \rangle$. Thus, if the number of satellite galaxies comes from a Poisson distribution, then $\big \langle N(N-1) \big \rangle_M / \big \langle N \big \rangle^2 $ should be equal to 1 \citep{2003ApJ...593....1B}. In Figure~\ref{fig:poisson} we have plotted this quantity for the Illustris (left, red) and EAGLE (right, blue) $M_r^{-19}$ samples as a function of halo mass. We have also plotted percentiles for our 300 HOD realizations for each sample (shown in gray), as well as the median of the 300 realizations. In our HOD model, the number of satellite galaxies is drawn from a Poisson distribution by design, so the median of these realizations should be 1 for all halo mass bins above $M_\mathrm{min}$ (indicated by the vertical green dashed line; below $M_\mathrm{min}$ it is extremely unlikely that there will be any satellites, so this quantity should be 0.) Both the Illustris and EAGLE samples are Poissonian at higher halo masses, but appear slightly sub-Poissonian at lower halo masses. However, neither sample is incompatible with its corresponding distribution of HOD realizations, so it is reasonable to conclude that the satellite numbers in Illustris and EAGLE are consistent with our HOD model. (The $M_r^{-21}$ samples have very few satellites, and thus are very noisy, which is why they are not shown here. They do not exhibit any non-Poissonian trends.)

\begin{figure*}
    \centering
	\includegraphics[width=\textwidth]{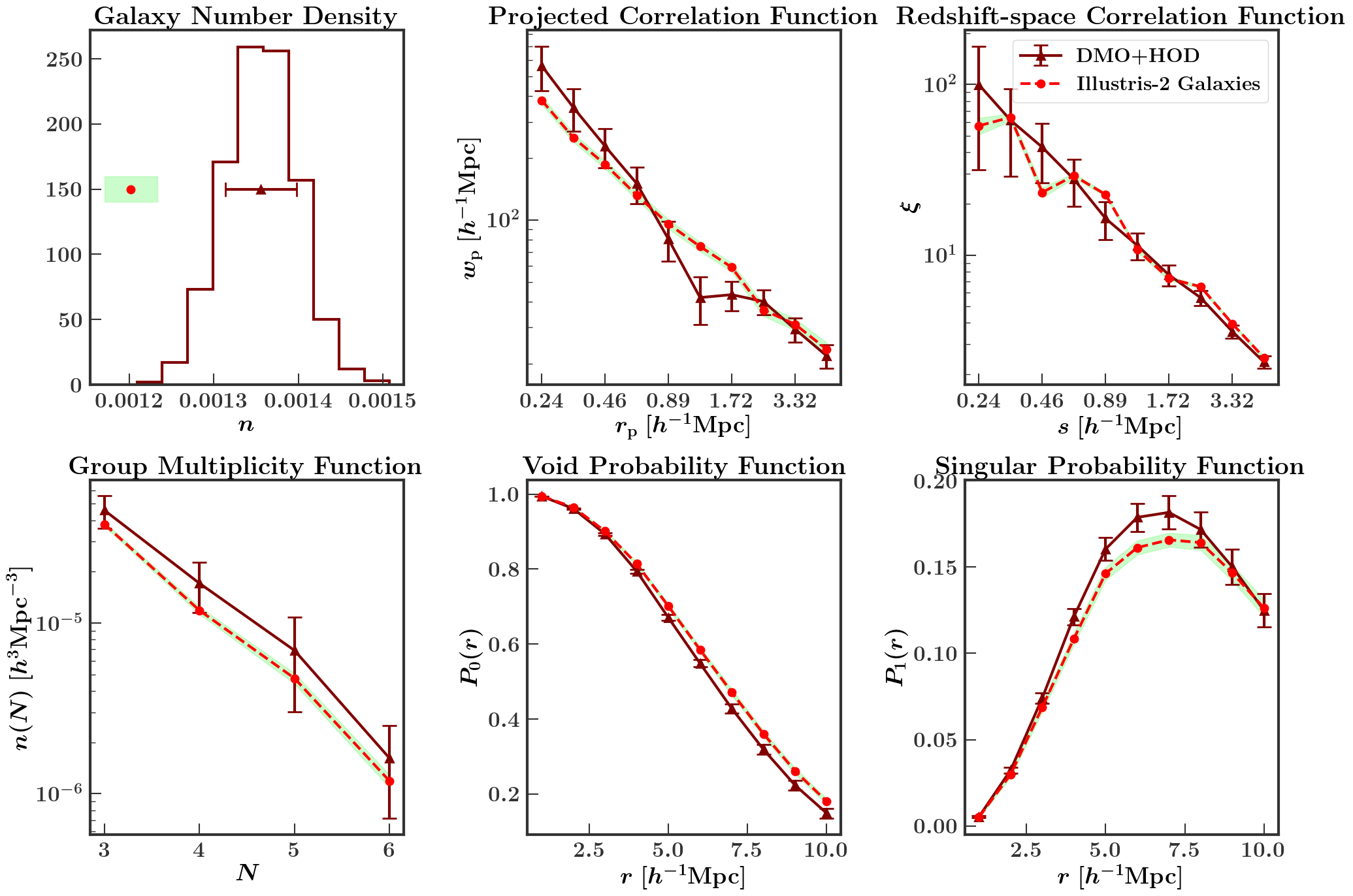}
    \caption{All clustering measurements for the $M_r^{-21}$ sample of Illustris-2 galaxies. The red lines are measured on galaxies from the original hydrodynamic simulation, while the dark red lines show the average of 1000 realizations of the best-fitting HOD model applied to the dark matter only simulation. The error bars represent the standard deviation among the 1000 realizations. The shaded regions around the red lines show cosmic variance errors (one standard deviation) calculated from 400 mock galaxy catalogues of the SDSS $M_r^{-21}$ sample, and thus illustrate the size of deviations that could be detected by the SDSS.}
    \label{fig:illustris_21_dark_all_stats}
\end{figure*}

\subsection{Building mock galaxy catalogs}
\label{sec:populating}
Once we have determined the best-fitting HOD parameters for our sample, we then need to actually place galaxies in haloes. We do this on the dark matter only versions of the simulations. As stated earlier, the halo mass of interest is the total mass of the Friends-of-Friends group (i.e. parent halo). We assign the central galaxy the position of the group, which is defined as the spatial position within the periodic box of the particle with the minimum gravitational potential energy (in comoving coordinates). Additionally, we assign the central galaxy the velocity of the group, which is the sum of the mass weighted velocities of all particles/cells in the group. The peculiar velocity is obtained by multiplying this value by $1/a$, where $a$ is the scale factor. (In the EAGLE simulation, the velocity of the parent halo is not provided, so we instead assign the central galaxy the velocity of the central subhalo.) To place satellite galaxies, we randomly select dark matter particles from the parent halo and assign galaxies the positions and velocities of these randomly chosen particles. The only stipulation we make is that we never choose the same random dark matter particle twice; i.e. we will never place two galaxies on the same particle, but we can place them on very nearby particles. We repeat this process 1000 times, so that we ultimately have 1000 different realizations of our best-fitting HOD model applied to our dark matter only simulation. We will refer to these 1000 realizations as mock galaxy catalogues.

\section{Galaxy Clustering Measurements} \label{stats}
Once we have populated the dark matter haloes in each simulation with galaxies, the next step is to measure a series of clustering statistics on both the galaxies from the original simulation and the galaxies from our mock catalogues. We measure these statistics in the same way on the simulation galaxies as we do on our mocks, in order to assess how well our HOD model can reproduce galaxy clustering properties as compared to a full hydrodynamic simulation.

\begin{figure*}
	\includegraphics[width=\textwidth]{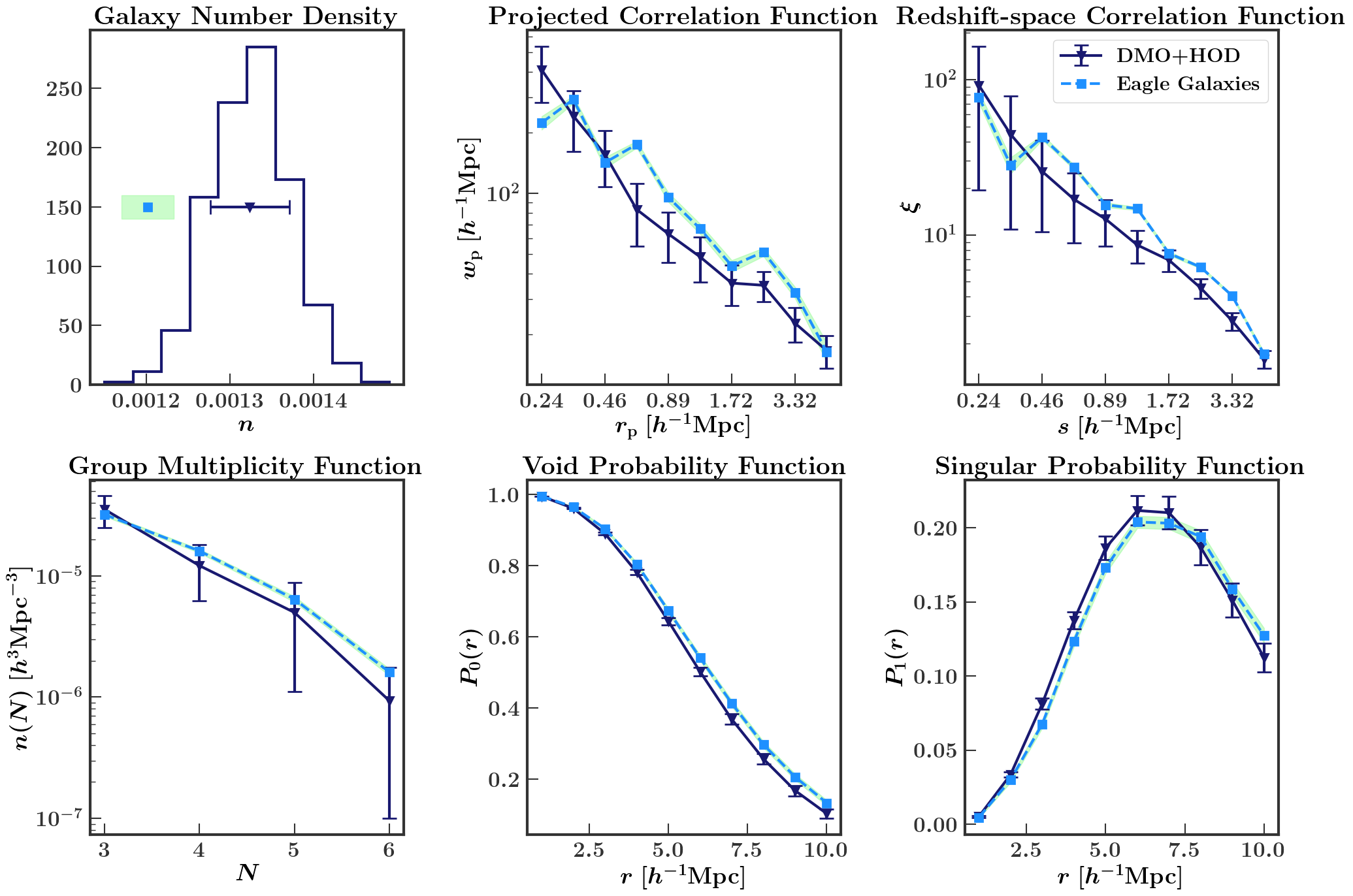}
    \caption{Same as Fig.~\ref{fig:illustris_21_dark_all_stats} for the $M_r^{-21}$ sample of EAGLE galaxies.}
    \label{fig:eagle_21_dark_all_stats}
\end{figure*}

\begin{figure*}
    \centering
	\includegraphics[width=\textwidth]{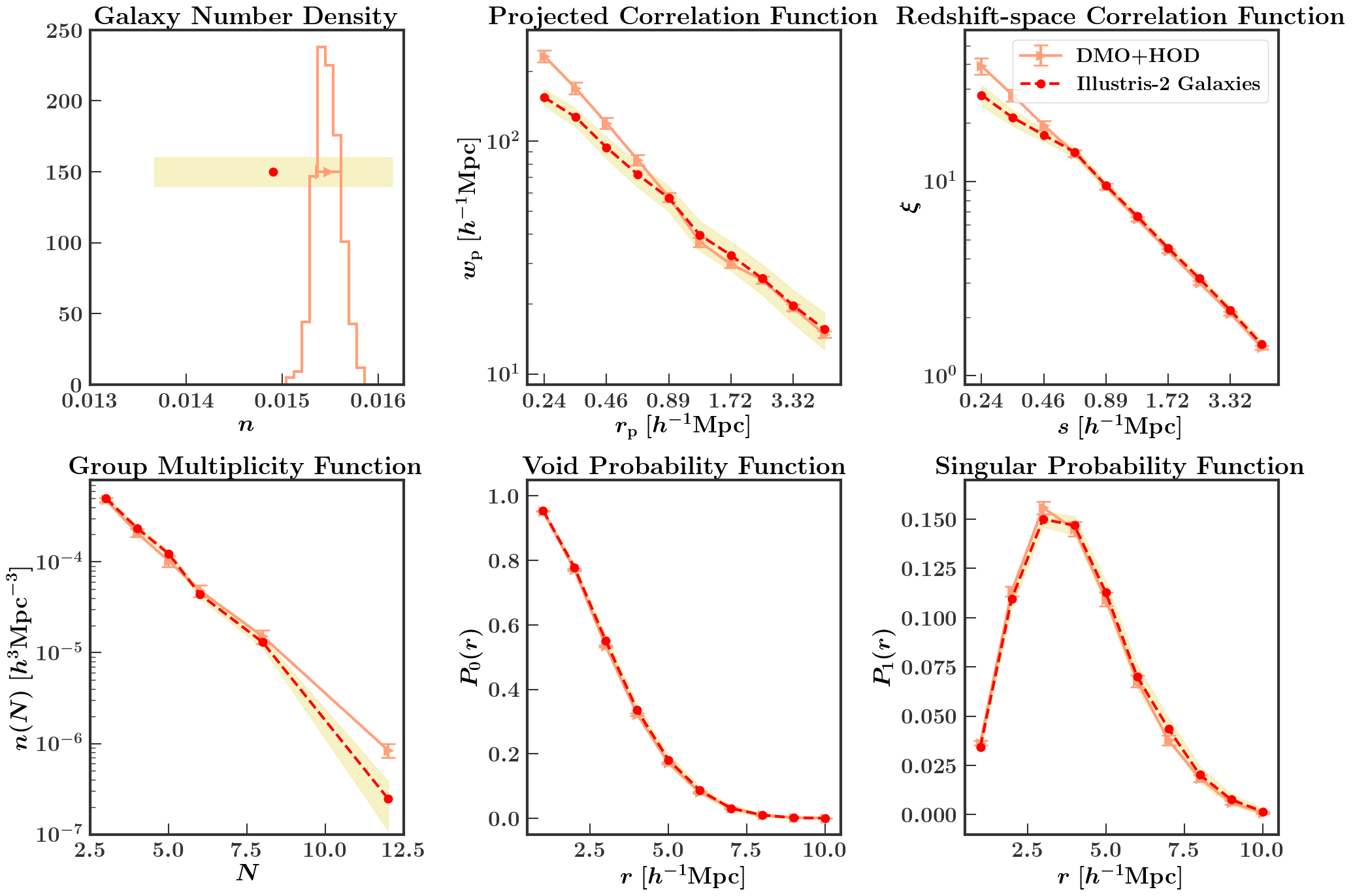}
    \caption{Same as Fig.~\ref{fig:illustris_21_dark_all_stats} for the $M_r^{-19}$ sample of Illustris-2 galaxies.}
    \label{fig:illustris_19_dark_all_stats}
\end{figure*}

The first property that we measure is the number density of galaxies. By comparing the number densities of galaxies in our simulations and in our mocks, we can test how well the HOD fits the simulation, as well as how similar the halo mass functions are in the hydrodynamic and dark matter only simulations. Figures~\ref{fig:illustris_21_dark_all_stats}-\ref{fig:eagle_19_dark_all_stats} show results for the Illustris $M_r^{-21}$, EAGLE $M_r^{-21}$, Illustris $M_r^{-19}$, and EAGLE $M_r^{-19}$ samples, respectively. The top left panel of each figure shows the distribution of number densities among the 1000 mocks for that sample (together with the mean and standard deviation), as well as the number density for the corresponding hydrodynamic sample. The shaded region in each figure shows cosmic variance errors (one standard deviation) calculated from 400 mock galaxy catalogues of the corresponding SDSS sample \citep{2018MNRAS.478.1042S}. The spread among our 1000 HOD mocks indicates how well we can measure galaxy number density in a box given the scatter in our HOD model. The spread among 400 SDSS mocks indicates how accurately a difference in number density could be detected by the SDSS.

In every case, applying the HOD to the dark matter only simulation results in a significantly overestimated galaxy number density (by up to 20\% for the Illustris $M_r^{-21}$ sample). For both $M_r^{-21}$ samples (Figures~\ref{fig:illustris_21_dark_all_stats} and \ref{fig:eagle_21_dark_all_stats}), this difference in number density is larger than the cosmic variance error from the SDSS $M_r^{-21}$ sample (shown in green); in other words, an SDSS-like survey would easily notice this discrepancy. For the $M_r^{-19}$ samples (Figures~\ref{fig:illustris_19_dark_all_stats} and \ref{fig:eagle_19_dark_all_stats}), although the difference between the simulation and the HOD number density is quite significant, the cosmic variance error (shown in yellow) is larger, indicating that an SDSS-like survey would not pick up on this difference. None the less, it is shocking that in every case the HOD (which was fit to the simulation) systematically significantly overestimates the galaxy number density. This points to a major issue with applying HOD to a dark matter only simulation: the halo mass function is different in hydrodynamic and dark matter only simulations. This will be discussed further in Section~\ref{hmf}.

Next, we measure five additional clustering statistics. Before we can do this, we must introduce redshift-space distortions into both our simulation galaxies as well as our mock galaxies. We do this by placing an observer infinitely far away from our box and taking the z-axis as the line of sight coordinate (using periodic boundary conditions). Including these distortions allows us to probe how well our model reproduces the velocities of the galaxies.

\citet{2002ApJ...575..587B} investigated galaxy bias in an HOD framework by measuring several clustering statistics. They found that the galaxy correlation function is affected by different parts of the HOD on different scales, and that other clustering statistics (such as the void probability function and the group multiplicity function) are also sensitive to different combinations of HOD parameters. \citet{2018MNRAS.478.1042S} similarly found that analyses involving several different galaxy clustering statistics have the most power to constrain galaxy bias. Because of this, the five additional clustering statistics that we measure in this work are the redshift-space correlation function, the projected correlation function, the group multiplicity function, the void probability function, and what we call the ``singular probability function" (i.e. the probability of having exactly one galaxy in a region). These five different clustering statistics are described in detail below.

\subsection{The projected correlation function}
\label{sec:wp}
The most commonly used galaxy clustering statistic, the projected correlation function, removes the effect of redshift-space distortions by first counting pairs of galaxies in bins of their line-of-sight and projected components, $\pi$ and $r_\mathrm{p}$, and then integrating over $\pi$:
\begin{equation}
w_\mathrm{p}(r_\mathrm{p}) = 2 \int_{0}^{\pi_\mathrm{max}}\xi(r_\mathrm{p},\pi)d\pi.
\end{equation}

We count pairs of galaxies in 10 evenly spaced logarithmic bins of projected separation $r_p$ between $0.2$ and $5.37 h^{-1} \mathrm{Mpc}$. We then integrate out to $\pi_\mathrm{max}$ of $20 h^{-1} \mathrm{Mpc}$ for each sample. (For computational reasons, $\pi_\mathrm{max}$ must be $< \frac{1}{3} L_{box}$.) We use the blazing fast code \texttt{Corrfunc} \citep{2017ascl.soft03003S,2019MNRAS.tmp.2750S} to compute our projected correlation function. 

The projected correlation function has been used as the workhorse of HOD modelling \citep[e.g.,][]{2011ApJ...736...59Z,2018MNRAS.478.1042S}. Recently, \citet{2019MNRAS.485.1196Z} used measurements of the projected correlation function to constrain assembly bias of SDSS DR7 galaxies within the decorated HOD model of \citet{2016MNRAS.460.2552H}. The authors found highly significant central galaxy assembly bias in the $M_r^{-20}$ and $M_r^{-20.5}$ samples, as well as significant satellite galaxy assembly bias for the $M_r^{-19}$ sample. They did not find any assembly bias in the $M_r^{-21}$ sample. Meanwhile, \citet{2019ApJ...872..115V} also looked at clustering measurements of SDSS DR7 galaxies and found that at fixed halo mass, satellite galaxies show no correlation with halo concentration, and central galaxies shows little correlation with halo concentration for the $M_r^{-21}$ and $M_r^{-21.5}$ samples, and slight correlation with halo concentration in the $M_r^{-20.5}$, $M_r^{-20}$, and $M_r^{-19}$ samples.

\begin{figure*}
	\includegraphics[width=\textwidth]{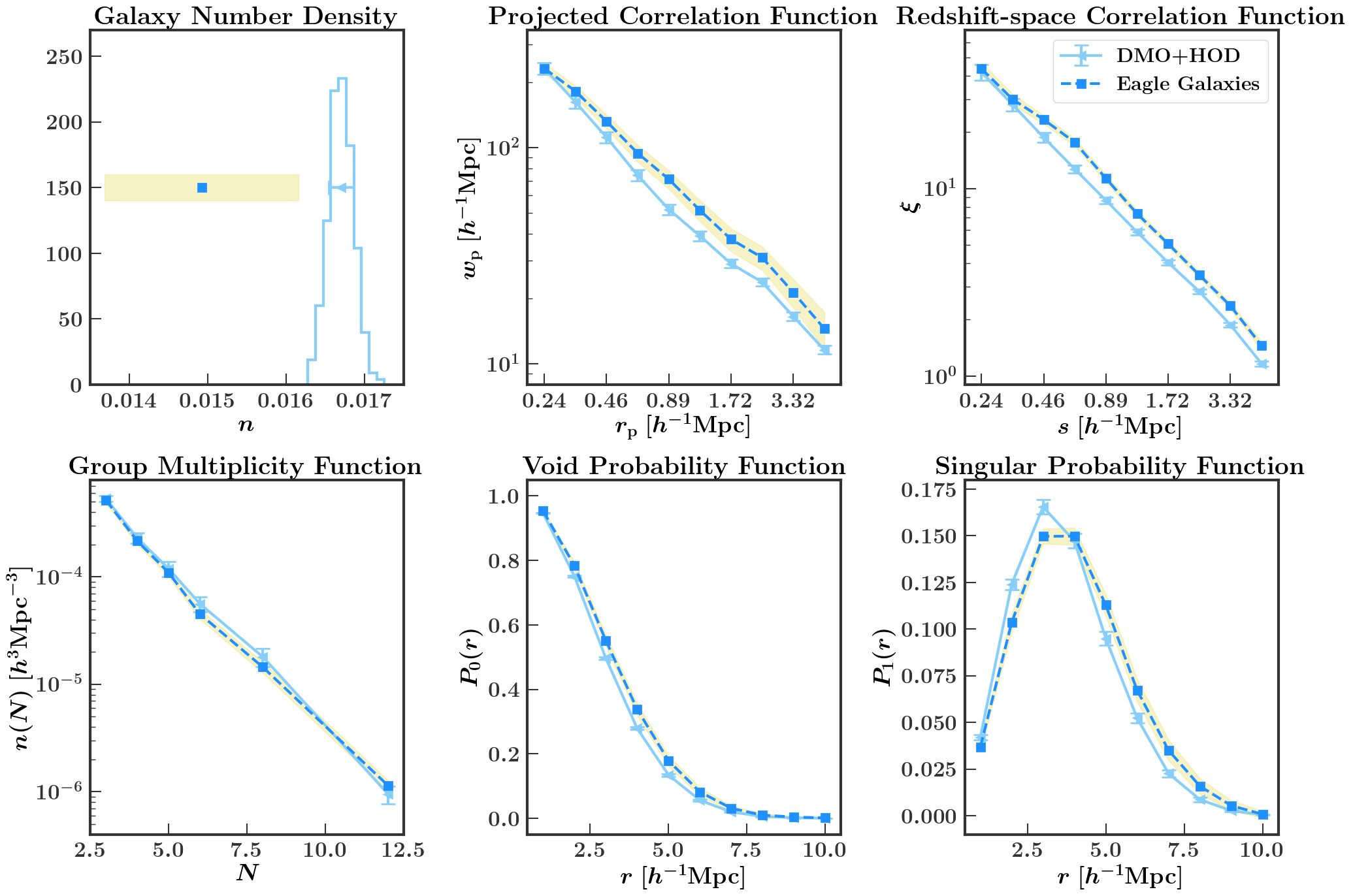}
    \caption{Same as Fig.~\ref{fig:illustris_21_dark_all_stats} for the $M_r^{-19}$ sample of EAGLE galaxies.}
    \label{fig:eagle_19_dark_all_stats}
\end{figure*}

In the top middle panels of Figures~\ref{fig:illustris_21_dark_all_stats}--\ref{fig:eagle_19_dark_all_stats} we have plotted the projected correlation function from the hydrodynamic simulations, as well as the average projected correlation function of our 1000 dark matter only mocks, for each of our samples. For the $M_r^{-21}$ samples (Figures~\ref{fig:illustris_21_dark_all_stats} and \ref{fig:eagle_21_dark_all_stats}) the HOD does reasonably well at recovering the projected correlation function from the simulations. Though there are visible discrepancies, these are not highly significant given the plotted uncertainties. However, for the Illustris $M_r^{-19}$ sample (Fig.~\ref{fig:illustris_19_dark_all_stats}), the HOD significantly overestimates the projected correlation function at small scales. In contrast, for the EAGLE $M_r^{-19}$ sample (Fig.~\ref{fig:eagle_19_dark_all_stats}), the HOD significantly underestimates the projected correlation function at all but the smallest scales. This indicates that although the clustering is correct for high luminosity galaxies, there is a possible problem with the spatial assumptions made in the HOD, which specifically impacts the clustering of low luminosity galaxies. The Illustris $M_r^{-19}$ sample is most likely affected by spatial bias, which impacts small scales, while the EAGLE $M_r^{-19}$ is likely more affected by assembly bias, which impacts large scales. We note that the projected correlation function is not sensitive to velocity information, so any discrepancies must be due to spatial and/or assembly bias, and not velocity bias. These biases will be discussed further in Section~\ref{decorated}. 

\subsection{The redshift-space correlation function}
\label{sec:xi}
The three-dimensional redshift-space two-point correlation function $\xi(s)$ is the excess number of galaxy pairs above that which is expected for a random distribution of points, as a function of redshift-space pair separation $s$ (in contrast to the projected separation $r_p$ described above). In this work, we count pairs in 10 bins of separation $s$ between $0.2$ and $5.37 h^{-1}\mathrm{Mpc}$ (the same bins as those used for the projected correlation function).  We also use \texttt{Corrfunc} to compute our redshift-space correlation function. Measuring the redshift-space correlation function allows us to access not only spatial information about our galaxies, but also velocity information, because the redshift-space distortions of our galaxies depend on their velocities. Thus, with this measurement, we can examine the validity of the assumption in the HOD that galaxies trace the velocity distribution of dark matter within the halo (in addition to examining our assumptions about the spatial distribution of galaxies).

In the top right panels of Figures~\ref{fig:illustris_21_dark_all_stats}--\ref{fig:eagle_19_dark_all_stats} we have plotted the redshift-space correlation function from our simulations, as well as the average redshift-space correlation function of our 1000 mocks, for each of our samples. Results are qualitatively similar to those using the projected correlation function. For the $M_r^{-21}$ samples (Figs~\ref{fig:illustris_21_dark_all_stats} and \ref{fig:eagle_21_dark_all_stats}) the HOD successfully recovers the redshift-space correlation function from the simulations. However, for the Illustris $M_r^{-19}$ sample (Fig.~\ref{fig:illustris_19_dark_all_stats}), the HOD once again significantly overestimates the correlation function at small scales, while for the EAGLE $M_r^{-19}$ sample (Fig.~\ref{fig:eagle_19_dark_all_stats}), the HOD significantly underestimates the correlation function at all but the smallest scales. This again suggests a problem with the spatial assumptions made in the HOD, as well as the velocity assumptions, which specifically impact the clustering of low luminosity galaxies. This will be discussed further in Section~\ref{decorated}.

\subsection{The group multiplicity function}
\label{sec:gmf}
The group multiplicity function is the abundance of galaxy groups as a function of the number of galaxies in the group, $n(N)$ \citep[e.g.,][]{2002ApJ...575..587B}. We use the \citet{2006ApJS..167....1B} friends-of-friends algorithm for identifying groups. Galaxies are linked together if their projected and line-of-sight separations are both less than a corresponding linking length. We adopt the \citet{2006ApJS..167....1B} linking lengths of $b_{\bot}=0.14$ and $b_{\parallel}=0.75$, which are given in units of the mean inter-galaxy separation $n_\mathrm{g}^{-1/3}$, where $n_\mathrm{g}$ is the sample number density. For our low luminosity samples, we measure groups with the following numbers of galaxies: $3, 4, 5, 6-7, 8-11, >12$. For our high luminosity samples, we measure groups of 3, 4, 5, and 6 or more galaxies.

In the lower left panels of Figures~\ref{fig:illustris_21_dark_all_stats}--\ref{fig:eagle_19_dark_all_stats} we have plotted the group multiplicity function from our simulations, as well as the average group multiplicity function of our 1000 mocks, for each of our samples. For the $M_r^{-21}$ samples (Figures~\ref{fig:illustris_21_dark_all_stats} and \ref{fig:eagle_21_dark_all_stats}) the HOD successfully recovers the group multiplicity function from the simulations. The HOD also successfully reproduces the group multiplicity function for the EAGLE $M_r^{-19}$ sample (Fig.~\ref{fig:eagle_19_dark_all_stats}). However, for the Illustris $M_r^{-19}$ sample (Fig.~\ref{fig:illustris_19_dark_all_stats}), the HOD significantly overestimates the group multiplicity function for the largest groups. This further points to a problem with the spatial and velocity assumptions made in the HOD, particularly as they affect the clustering of low luminosity galaxies in Illustris. This will be discussed further in Section~\ref{decorated}.

\subsection{Counts-in-cells statistics}
\label{sec:cic}
Counts-in-cells statistics measure the probability of finding a given number of galaxies within a randomly placed finite region (e.g. a sphere) as a function of region size (e.g. radius). One special case of this is the void probability function (VPF), which measures the probability of finding no galaxies in a random region of space. \citet{2006ApJ...647..737T} attempted to constrain galaxy bias using void statistics within an HOD framework, and found that the VPF, in contrast to the projected correlation function, is quite sensitive to environmental variations of the HOD. Later, \citet{2017arXiv170501988M} showed that catalogues created using SHAM and the semi-analytic model \texttt{galform}, which were designed to have the same large-scale 2-point clustering, have different VPFs due to their different HOD shapes, suggesting that the VPF could be used to rule out certain HOD models. Recently, \citet{2019MNRAS.tmp.1286W} fit the standard HOD model to the two-point correlation function of BOSS galaxies and found that it was able to accurately predict the void probability function, indicating that galaxy assembly bias does not affect the clustering of massive galaxies.

\citet{2019MNRAS.tmp.1686W} studied the power of the VPF, counts-in-cylinders, and counts-in-annuli, as well as the projected two-point correlation function and the galaxy-galaxy lensing signal to constrain galaxy assembly bias from redshift survey data using the decorated HOD, and found that the counts-in-cells statistics are more efficient at constraining galaxy assembly bias when combined with the projected correlation function than galaxy-galaxy lensing is.

Another variation of counts in cells that we use is what we will refer to as the ``singular probability function," (SPF) or the probability of finding exactly one galaxy in a randomly placed region. We measure both the VPF and the SPF in spheres of evenly spaced bins of radius $r$, beginning with $1 h^{-1}\mathrm{Mpc}$ and ending with $10 h^{-1}\mathrm{Mpc}$.

\begin{table*}
\centering
\caption{$p$-values from comparing the clustering statistics of hydrodynamic galaxies to those of DMO+HOD mock galaxies, for 
different simulations and samples, with no correction (first), after correcting the halo mass function (second), additionally removing satellite spatial bias (third), additionally removing all spatial and velocity bias (fourth), and additionally removing assembly bias (fifth). The columns show (from left to right): simulation name, magnitude limit for the SDSS sample with the same galaxy number density, which model was used, and the $p$-values for each of our six measurements.}
\label{tab:all_pval_table}
\begin{tabular}{cccccccccc}
\hline
Sim. & Sample & Correction & $n$ & $w_p(r_p)$ & $\xi(s)$ & $n(N)$ & $P_0(r)$ & $P_1(r)$\\
\hline 
Illustris & -21 & No Correction & $2.84 \times 10^{-4}$ & $4.61 \times 10^{-2}$ & $6.62 \times 10^{-1}$ & $5.36 \times 10^{-1}$ & $1.86 \times 10^{-2}$ & $4.04 \times 10^{-1}$\\
Illustris & -21 & Halo Mass Function & $4.54 \times 10^{-1}$ & $1.43 \times 10^{-1}$ & $9.14 \times 10^{-1}$ & $9.77 \times 10^{-1}$ & $4.79 \times 10^{-1}$ & $6.28 \times 10^{-1}$\\
Illustris & -21 & +Satellite Spatial Bias & $4.54 \times 10^{-1}$ & $6.46 \times 10^{-1}$ & $7.51 \times 10^{-1}$ & $6.95 \times 10^{-1}$ & $4.84 \times 10^{-1}$ & $6.35 \times 10^{-1}$\\
Illustris & -21 & +Velocity Bias & $4.54 \times 10^{-1}$ & $6.66 \times 10^{-1}$ & $5.98 \times 10^{-1}$ & $7.09 \times 10^{-1}$ & $3.97 \times 10^{-1}$ & $6.13 \times 10^{-1}$\\	
Illustris & -21 & +Assembly Bias & $4.54 \times 10^{-1}$ & $6.15 \times 10^{-1}$ & $5.25 \times 10^{-1}$ & $6.11 \times 10^{-1}$ & $5.23 \times 10^{-1}$ & $6.87 \times 10^{-1}$\\
\hline
Illustris & -19 & No Correction & $8.35 \times 10^{-6}$ & $1.13 \times 10^{-7}$ & $1.61 \times 10^{-4}$ & $5.36 \times 10^{-4}$ & $4.44 \times 10^{-6}$ & $5.99 \times 10^{-2}$\\
Illustris & -19 & Halo Mass Function & $6.66 \times 10^{-2}$ & $5.23 \times 10^{-6}$ & $2.43 \times 10^{-3}$ & $3.48 \times 10^{-4}$ & $1.05 \times 10^{-3}$ & $5.25 \times 10^{-2}$\\
Illustris & -19 & +Satellite Spatial Bias & $6.66 \times 10^{-2}$ & $2.58 \times 10^{-2}$ & $1.14 \times 10^{-1}$ & $1.87 \times 10^{-2}$ & $2.11 \times 10^{-3}$ & $5.69 \times 10^{-2}$ \\
Illustris & -19 & +Velocity Bias & $6.66 \times 10^{-2}$ & $2.89 \times 10^{-2}$ & $1.94 \times 10^{-1}$ & $8.76 \times 10^{-2}$ & $9.68 \times 10^{-2}$ & $4.42 \times 10^{-1}$ \\	
Illustris & -19 & +Assembly Bias & $6.66 \times 10^{-2}$ & $7.64 \times 10^{-2}$ & $4.81 \times 10^{-1}$ & $1.65 \times 10^{-1}$ & $3.93 \times 10^{-1}$ & $7.82 \times 10^{-1}$\\
\hline
EAGLE & -21 & No Correction & $9.84 \times 10^{-3}$ & $5.89 \times 10^{-3}$ & $3.69 \times 10^{-3}$ & $8.18 \times 10^{-1}$ & $5.32 \times 10^{-2}$ & $1.91 \times 10^{-2}$\\
EAGLE & -21 & Halo Mass Function & $8.56 \times 10^{-1}$ & $3.64 \times 10^{-2}$ & $4.07 \times 10^{-2}$ & $7.02 \times 10^{-1}$ & $5.55 \times 10^{-1}$ & $2.01 \times 10^{-1}$\\
EAGLE & -21 & +Satellite Spatial Bias & $8.56 \times 10^{-1}$ & $4.05 \times 10^{-1}$ & $1.53 \times 10^{-1}$ & $9.18 \times 10^{-2}$ & $6.99 \times 10^{-1}$ & $2.92 \times 10^{-1}$\\
EAGLE & -21 & +Velocity Bias & $8.56 \times 10^{-1}$ & $4.06 \times 10^{-1}$ & $2.53 \times 10^{-1}$ & $1.98 \times 10^{-1}$ & $6.61 \times 10^{-1}$ & $2.69 \times 10^{-1}$ \\
EAGLE & -21 & +Assembly Bias & $8.56 \times 10^{-1}$ & $3.08 \times 10^{-1}$ & $5.55 \times 10^{-1}$ & $4.84 \times 10^{-1}$ & $3.55 \times 10^{-1}$ & $4.06 \times 10^{-1}$\\
\hline
EAGLE & -19 & No Correction & $6.37 \times 10^{-29}$ & $1.11 \times 10^{-13}$ & $1.63 \times 10^{-24}$ & $4.50 \times 10^{-1}$ & $7.11 \times 10^{-54}$ & $3.37 \times 10^{-22}$\\
EAGLE & -19 & Halo Mass Function & $8.25 \times 10^{-1}$ & $1.06 \times 10^{-8}$ & $3.42 \times 10^{-10}$ & $6.31 \times 10^{-1}$ & $4.79 \times 10^{-13}$ & $1.42 \times 10^{-7}$\\
EAGLE & -19 & +Satellite Spatial Bias & $8.25 \times 10^{-1}$ & $3.90 \times 10^{-5}$ & $2.22 \times 10^{-8}$ & $1.13 \times 10^{-1}$ & $8.58 \times 10^{-13}$ & $1.87 \times 10^{-7}$ \\
EAGLE & -19 & +Velocity Bias & $8.25 \times 10^{-1}$ & $6.80 \times 10^{-5}$ & $2.40 \times 10^{-5}$ & $2.24 \times 10^{-1}$ & $7.90 \times 10^{-10}$ & $6.50 \times 10^{-5}$ \\
EAGLE & -19 & +Assembly Bias & $8.25 \times 10^{-1}$ & $1.49 \times 10^{-1}$ & $3.10 \times 10^{-1}$ & $4.92 \times 10^{-1}$ & $4.97 \times 10^{-1}$ & $6.07 \times 10^{-1}$\\		
\hline
\end{tabular}
\end{table*}

In the lower middle (right) panels of Figures~\ref{fig:illustris_21_dark_all_stats}--\ref{fig:eagle_19_dark_all_stats} we have plotted the VPF (SPF) of our simulations, as well as the average of our 1000 mocks, for each of our samples. For the Illustris $M_r^{-21}$ sample (Fig.~\ref{fig:illustris_21_dark_all_stats}) the HOD struggles to recover the VPF at intermediate and large scales, and likewise struggles to recover the SPF at intermediate scales. For the EAGLE $M_r^{-21}$ sample (Fig.~\ref{fig:eagle_21_dark_all_stats}) the HOD shows similar tension in the VPF and the SPF. For the Illustris $M_r^{-19}$ sample the agreement looks better, but the error bars are very small so it is difficult to surmise based on looking at Figure~\ref{fig:illustris_19_dark_all_stats} alone. For the EAGLE $M_r^{-19}$ sample (Fig.~\ref{fig:eagle_19_dark_all_stats}) the HOD struggles to reproduce both the VPF and the SPF at most scales. These problems could indicate issues with the assumptions made in the HOD. They could also be compounded by the inability of the HOD to reproduce the correct number density, since counts-in-cells statistics, and the VPF in particular, are very sensitive to number density. This will be discussed further in Section~\ref{decorated}.

\section{Assessing the Accuracy of the HOD Model} \label{fit}
In Figures~\ref{fig:illustris_21_dark_all_stats}--\ref{fig:eagle_19_dark_all_stats} we saw that for some statistics (like number density) the HOD applied to dark matter only simulations does not provide a good fit to the hydrodynamic simulations for any of our samples, while for other statistics (like the correlation functions) the HOD appeared to provide a good fit to the simulations for the high luminosity samples and not the low luminosity samples. In general, however, the success of the HOD model is difficult to ascertain visually because error-bars are often small and are likely correlated. In order to quantify the accuracy with which our HOD model can reproduce the clustering statistics measured on a hydrodynamic simulation, we calculate $\chi^2$ for each clustering statistic
\begin{equation}
\chi^2 = \sum_{ij} \chi_i R_{ij}^{-1}\chi_j,
\end{equation}
where
\begin{equation}
\chi_i = \frac{D_i - M_i}{\sigma_i},
\end{equation}
$D_i$ is the value of one bin of a clustering measurement on the hydrodynamic simulation galaxies (either Illustris or EAGLE, and either $M_r^{-19}$ or $M_r^{-21}$), $M_i$ is that same measurement averaged over our 1000 mock galaxy catalogues for that sample, and $\sigma_i$ is the standard deviation of that measurement among the 1000 mock galaxy catalogues. $R_{ij}$ is the correlation matrix for each clustering statistic
\begin{equation}
R_{ij} = \frac{C_{ij}}{\sqrt{C_{ii}C_{jj}}},
\end{equation}
which is the covariance matrix normalized by its diagonal elements. The covariance matrix is calculated as
\begin{equation}
C_{ij} = \frac{1}{N-1} \sum_{1}^{N} (y_i - \overline{y_i})(y_j - \overline{y_j}),
\end{equation}
where the sum is over the $N = 1000$ mock galaxy catalogues, and $y_i$ and $y_j$ are two bins of a clustering statistic, and $\overline{y_i}$ and $\overline{y_j}$ are the mean measurements over the 1000 mocks. We note that since the hydrodynamic simulation and the HOD mocks come from initial conditions with the same phases, cosmic variance errors do not apply to this comparison.

From this $\chi^2$, we can calculate the corresponding $p$-value, which represents the probability that a sample randomly drawn from the best-fitting HOD model could have a $\chi^2$ value greater than the one exhibited by the simulation. In other words, the $p$-value represents the probability that the hydrodynamic simulation is consistent with the DMO+HOD model. The $p$-value for each clustering measurement uses all the spatial bins of the measurement, as well as the full covariance matrix for that statistic. These $p$-values are listed in Table~\ref{tab:all_pval_table} (in the rows labeled as ``No Correction''). 

Looking at Figures~\ref{fig:illustris_21_dark_all_stats}--\ref{fig:eagle_19_dark_all_stats} or the $p$-values in Table~\ref{tab:all_pval_table}, it is immediately clear that the vanilla HOD model, when applied to haloes from a dark matter only simulation, does not provide a good fit to the corresponding hydrodynamic simulation for all of the clustering statistics in question. However, the success of the HOD model is highly dependent on the simulation and luminosity sample in question. For example, the model generally performs better for high luminosity galaxies than for low luminosity galaxies. Specifically, for the Illustris $M_r^{-21}$ sample, all of the clustering statistics are well fit by the HOD model, at least within a $3\sigma$ tolerance, except for number density. For the EAGLE $M_r^{-21}$ sample, even the number density works well. However, for the low luminosity samples, almost none of the clustering statistics are well fit by the DMO+HOD model, and in most cases exhibit discrepancies far greater than $>3\sigma$.

The green shaded regions in Figures~\ref{fig:illustris_21_dark_all_stats} and \ref{fig:eagle_21_dark_all_stats} represent one standard deviation of cosmic variance errors calculated from 400 mock galaxy catalogues of the SDSS $M_r^{-21}$ sample. These mocks were created as part of the Large Suite of Dark Matter Simulations project \citep[LasDamas;][]{2009AAS...21342506M} and used in \citet{2018MNRAS.478.1042S}. In our $M_r^{-21}$ Illustris and EAGLE samples, the errors among our 1000 mock galaxy catalogues (which are different HOD realizations) are much larger than the cosmic variance errors from the 400 SDSS-like mocks. Consequently, though the HOD model appears to be a good fit to the simulations for high luminosity galaxies, an SDSS size $M_r^{-21}$ survey (which has small errors due to its large volume) could be sensitive to clustering differences that we are unable to detect in our analysis due to our smaller volume. 

Similarly, the yellow shaded regions in Figures~\ref{fig:illustris_19_dark_all_stats} and \ref{fig:eagle_19_dark_all_stats} represent one standard deviation of cosmic variance errors calculated from 400 mock galaxy catalogues of the SDSS $M_r^{-19}$ sample, constructed in a similar way as those in \citet{2018MNRAS.478.1042S}. In our $M_r^{-19}$ Illustris and EAGLE samples, the errors among our 1000 mock galaxy catalogues are smaller than the cosmic variance errors from the 400 SDSS-like mocks. For some statistics (such as the number density), a survey with the precision of SDSS would not necessarily be able to detect the differences we have found between the HOD model and the hydrodynamic simulation. For other clustering statistics (particularly the correlation functions) it is clear that, although the cosmic variance errors are somewhat broad, there is still an obvious difference between the HOD model and the simulation, to which even an SDSS-like survey would be sensitive.

\section{The Effect of Baryons on the Halo Mass Function} \label{hmf}

Figures~\ref{fig:illustris_21_dark_all_stats} -- \ref{fig:eagle_19_dark_all_stats} revealed that the galaxy number density is not well predicted in any sample. Recall that, in our vanilla HOD, the number of galaxies in a halo is solely dependent on the mass of the halo. Thus, the fact that our HOD systematically over-predicts the galaxy abundance indicates either that the functional form of our HOD is incorrect, or that the halo mass functions (HMFs) are different in the hydrodynamic simulations compared to their dark matter only (DMO) counterparts.

\begin{figure}
	\includegraphics[width=\columnwidth]{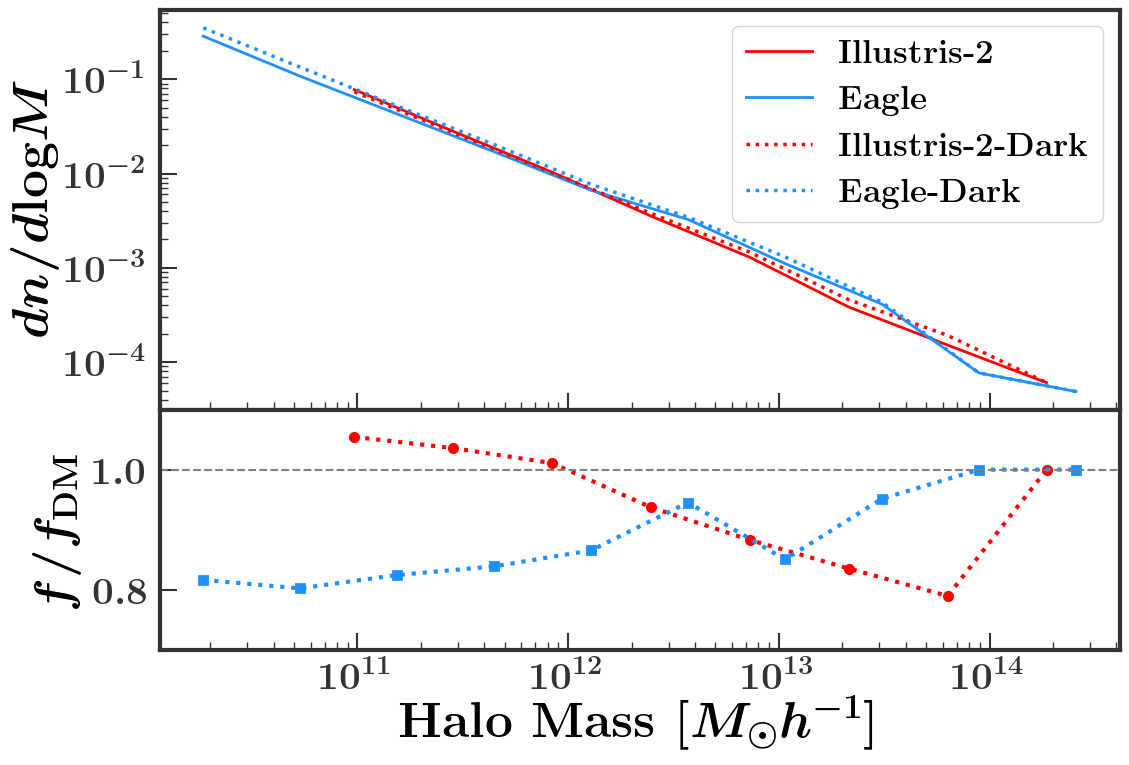}
    \caption{Halo mass functions of hydrodynamic compared to dark matter only simulations in the case of Illustris-2 (red) and EAGLE (blue). The hydrodynamic versions are plotted with solid lines, while the dark matter only versions are plotted with dotted lines. The bottom panel shows the ratio of the hydrodynamic to dark matter only mass functions for the two simulations.}
    \label{fig:illustris_eagle_hmf}
\end{figure}

Figure~\ref{fig:illustris_eagle_hmf} compares the abundance of haloes in the hydrodynamic and DMO versions of the same simulation. The comparison reveals sizeable discrepancies between the halo mass functions. In Illustris (red), the hydrodynamic HMF is consistently lower than the DMO HMF above $10^{12} h^{-1} M_\odot$, and higher than the DMO HMF at smaller masses. In EAGLE (blue), the hydrodynamic HMF is below the DMO HMF at all halo masses below $10^{14} h^{-1} M_\odot$. In other words, the hydrodynamic HMFs are shifted to lower masses in both simulations, but the detailed effects of baryons on the HMF are different in the two simulations.

This result is consistent with both \citet{2017MNRAS.471L..11D} and \citet{2015MNRAS.451.1247S}, who examined the differences between the halo masses in the EAGLE dark matter only and hydrodynamic runs, and found the haloes to be less massive on average in the hydrodynamic run. \citet{2017MNRAS.471L..11D} found that, at low halo masses, stellar feedback in EAGLE removes baryons from the halo, which in turn reduces the growth rate of the halo. At slightly higher halo masses, stellar feedback becomes less effective, but AGN feedback is still capable of expelling baryons. For the most massive haloes, AGN feedback too becomes less effective, and thus there is little discrepancy between the hydrodynamic and DMO halo mass functions.

Our results for the Illustris haloes are consistent with the findings of \citet{2014MNRAS.444.1518V}, who found that the halo mass function in Illustris is most affected at low ($< 10^{10} h^{-1} M_\odot$) and high ($> 10^{12} h^{-1} M_\odot$) halo masses, where baryonic feedback processes (e.g. reionization, SN feedback, and AGN feedback) are strongest, leading to a reduction in halo mass compared to their DMO counterparts. They found that removing AGN feedback boosts the massive end of the halo mass function \citep[e.g.][]{2012MNRAS.423.2279C}. They also found that haloes around $10^{11} h^{-1} M_\odot$, where star formation is most efficient, tend to be more massive than their DMO counterparts.

\begin{figure}
	\includegraphics[width=\columnwidth]{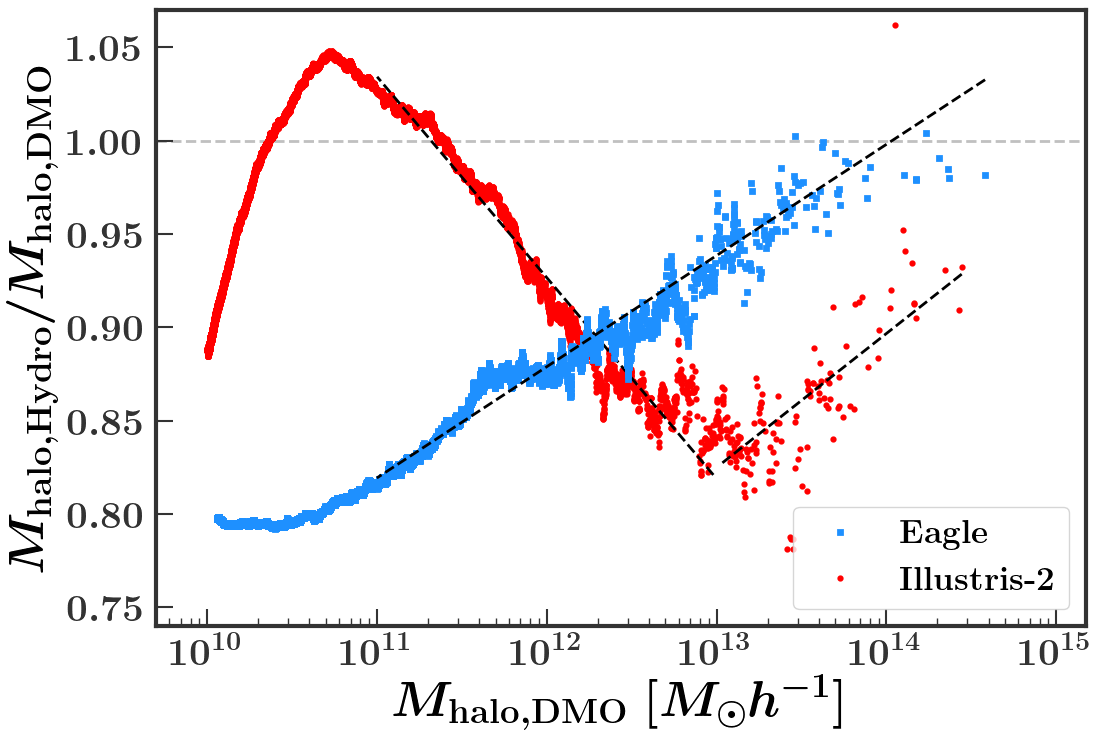}
    \caption{The ratio of halo masses from the hydrodynamic simulations to halo masses from the dark matter only simulations, as a function of dark matter only halo mass. Illustris-2 haloes are plotted in red and EAGLE haloes are plotted in blue. The halo mass is the total FoF mass from all particles, which in the hydrodynamic versions includes baryons. Hydrodynamic and dark matter only haloes are matched by their mass rank, rather than by position. The displayed ratio thus represents the correction factor needed to apply to the dark matter only haloes in order to recover the hydrodynamic mass function. The dashed black lines show simple fits to these relationships, down to $10^{11} h^{-1} M_\odot$, which we discuss in Section~\ref{summary}.}
    \label{fig:halo_masses}
\end{figure}

In Figure~\ref{fig:halo_masses} we show the ratio of halo masses in the hydrodynamic simulation over the masses in the DMO simulation as a function of halo mass in the DMO simulation, for both the Illustris-2 (red) and the EAGLE (blue) simulations. The hydrodynamic and DMO haloes are matched based on their ranked masses, rather than spatial positions, so that the point furthest to the right in the figure corresponds to the highest mass DMO halo, paired with the highest mass hydrodynamic halo. In other words, we essentially abundance match the haloes in the hydrodynamic and DMO simulations. As a result, the figure shows the mass correction one would need to apply to the DMO masses in order to recover the global hydrodynamic HMF. However, applying this correction would not necessarily result in the correct dependence of the HMF on environment.

Our result is consistent with the results of \citet{2014MNRAS.444.1518V} and \citet{2015MNRAS.451.1247S}, who looked at matched haloes in Illustris and EAGLE, respectively. Additionally, \citet{2018MNRAS.475..676S} looked at this same quantity for the IllustrisTNG simulations and found a trend that is different from both Illustris and EAGLE. Baryons in the IllustrisTNG seem to have a larger impact on low mass haloes and a smaller impact on high mass haloes compared to Illustris. This is to be expected, since IllustrisTNG has weaker AGN feedback than the original Illustris simulation, which affects more massive haloes. The effect of feedback on lower mass haloes in TNG is stronger than that in Illustris due to the wind model used in TNG.

Figure~\ref{fig:halo_masses} emphasizes the fact that the effect of baryons on the halo mass function is to decrease the HMF to lower masses. However, it is clear that this effect is very different in these two different simulations. The effect of baryons on the HMF in the EAGLE simulation is more prominent at lower masses, and the ratio of hydrodynamic halo mass to DMO halo mass increases almost linearly with log halo mass. In Illustris, the effect of baryons on the HMF is more prominent at higher masses, and the relationship is more complex than it is in EAGLE. In other words, the halo mass function is significantly affected by baryonic feedback processes, but there is no consensus among hydrodynamic simulations on what the correct feedback model is.

This halo mass function discrepancy presents a challenge when using an HOD framework to populate haloes from a dark matter only simulation with galaxies. The HOD parameters only describe how many galaxies to put in a halo of a given mass, but do not take into account how many haloes there are in a given mass bin. Therefore, because the dark matter only versions of Illustris and EAGLE have mass functions that are shifted to higher masses, there are more high mass haloes, so more galaxies are placed overall. Thus, even when applying the correct HOD parameters as extracted from the hydrodynamic simulation, the overall galaxy number density will be too high when this HOD is applied to the dark matter only simulation.

One possible solution to this is to adjust the HMF in the dark matter only simulation so that it is consistent with the HMF in the hydrodynamic version. We do this by identifying the most massive halo in the dark matter only simulation and assigning it the mass of the most massive halo in the hydrodynamic version, and then we do the same for the next most massive halo, and so on. In other words, we multiply the DMO halo masses in each simulation by their y-axis value in Figure~\ref{fig:halo_masses}. This process serves to isolate the effect of baryons on the halo mass function, allowing us to correct the DMO HMFs so that they agree with the HMFs from the hydrodynamic simulations. We note that this technique does not involve matching haloes based on position or particle-IDs. Because of this, we are not explicitly taking environment into account, so we are not correcting the conditional HMF. We have examined the conditional HMF in Illustris, however, and have found that the effect of baryons on the HMF only depends on environment at very high halo masses. Additionally, we have examined the effect on our clustering statistics if we use an environment-dependent HMF correction and find that the difference is negligible. We have also examined the halo correlation functions in Illustris and EAGLE in two different halo mass bins for the hydrodynamic simulations, the DMO simulations, and the corrected DMO simulations, and have found that the corrected DMO halo correlation functions are in better agreement with the hydrodynamic halo correlation functions.

\begin{figure*}
    \centering
    \includegraphics[width=6in]{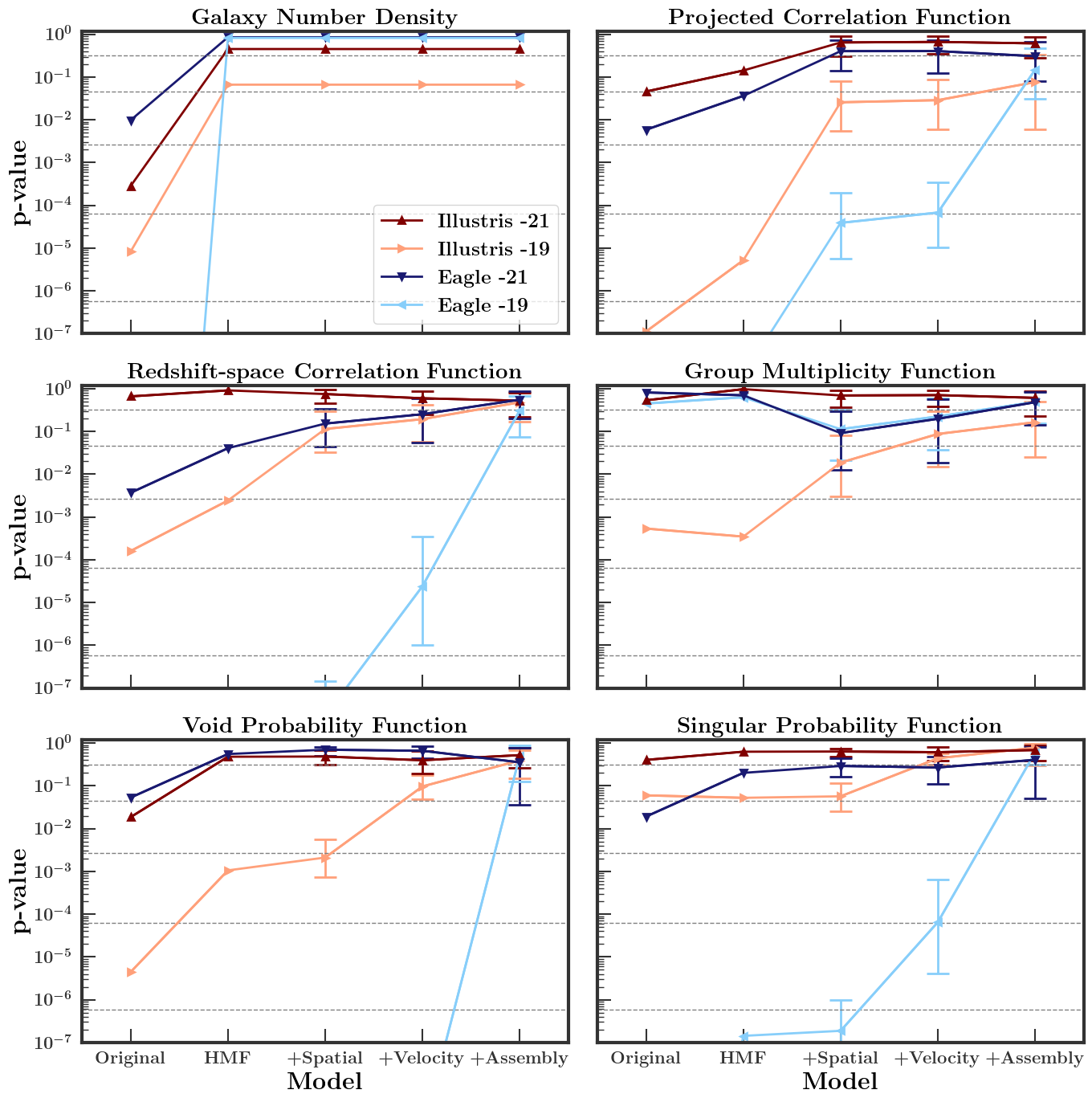}
    \caption{$p$-values from comparing the clustering of galaxies in hydrodynamic simulations to the clustering of mock galaxies in their dark matter only (DMO) counterparts. Each panel shows results for a different clustering statistic, as listed at the top of each panel. The dark red diamonds and dark blue squares represent the high luminosity samples of Illustris-2 and EAGLE, respectively, while the light red inverted triangles and the light blue triangles represent the low luminosity samples of Illustris-2 and EAGLE, respectively. The horizontal dashed gray lines denote the $1\sigma$, $2\sigma$, $3\sigma$, $4\sigma$, and $5\sigma$ confidence levels. The x-axis in each panel corresponds to different modifications to the haloes or to the galaxies in the simulations. From left to right, $p$-values are shown for (i) the original DMO+HOD model; (ii) the same DMO+HOD model after adjusting the DMO halo mass function to match the mass function in the hydrodynamic simulation; (iii) additionally removing satellite spatial bias from the hydrodynamic simulation galaxies; (iv) additionally removing central and satellite velocity bias from the hydrodynamic simulation galaxies; (v) additionally removing assembly bias from the hydrodynamic simulation galaxies. The last three $p$-values in each panel (with the exception of number density) are the median of many realizations (1000, 1000, and 4000), with error bars showing the 16th and 84th percentiles. For the low luminosity sample of EAGLE (light blue), several points are not shown because they fall below $10^{-7}$. The values of these points are given in Tables~\ref{tab:all_pval_table}.}
    \label{fig:pvals}
\end{figure*}

We now explore to what extent applying mass corrections to DMO halo masses improves the agreement between the clustering statistics of hydrodynamic and DMO+HOD galaxies. We first multiply each DMO halo mass by the correction shown in Figure~\ref{fig:halo_masses} (i.e. we use our abundance matching technique for each halo as described above, and not the dashed-black fits shown in the figure).  We then make new mock galaxy catalogues by applying the same best-fitting HOD (from Table~\ref{tab:hod_table}) to our new mass-adjusted dark matter haloes. We thus have 1000 new mock catalogues for each sample. We then repeat the same procedure outlined in Sections~\ref{stats} and \ref{fit} to get new clustering statistics and new $p$-values, which we list in Table~\ref{tab:all_pval_table} (in the rows labeled ``Halo Mass Function'').

Figure~\ref{fig:pvals} presents our $p$-values for the four samples (two simulations and two luminosity samples) for all six statistics we consider. The left-most point in each panel shows the original $p$-value we obtained and discussed in Section~\ref{fit}. The second point in each panel shows the new $p$-value we get after first applying a correction factor to the DMO halo masses. Horizontal dashed lines show the $1\sigma$, $2\sigma$, $3\sigma$, $4\sigma$, and $5\sigma$ tolerance levels. As we can see in Figure~\ref{fig:pvals}, after correcting the masses of haloes, our ability to accurately predict galaxy number density (top left panel) with our vanilla HOD model shows a drastic improvement for all samples. Thus, the vanilla form of HOD that we have adopted is sufficient for accurately (better than $2\sigma$ tolerance) predicting galaxy number density if it is applied to the correct population of haloes.

In addition to the improvement in our galaxy number density predictions for all samples, correcting the halo mass function yields a slight improvement to the other clustering statistics across all samples. For the $M_r^{-21}$ samples, after correcting the halo mass function, all clustering statistics are at or better than the $2\sigma$ level. Thus, when applied to the correct halo population, the 5 parameter HOD model is able to accurately predict all clustering statistics for our high luminosity samples of galaxies. For the low luminosity samples, although the other clustering statistics do improve, most are still below the $3\sigma$ level, with the exception of the group multiplicity function in the EAGLE $M_r^{-19}$ sample and the singular probability function in the Illustris $M_r^{-19}$ sample. It is worth noting that the VPF does improve in all samples after correcting the halo mass function, indicating that part of the original VPF discrepancy was due to the incorrect number density. However, for the Illustris $M_r^{-19}$ sample the VPF is still below the $3\sigma$ level, and for the EAGLE $M_r^{-19}$ sample it is still well below $5\sigma$, so we can conclude that not all of the issues with reproducing the VPF can be attributed to the number density.

These results indicate that although the HOD model for the brightest galaxies is successful when applied to the correct halo population, the HOD model for fainter galaxies is less successful, even when applied to the correct halo population. Thus, there must be some other assumptions in our HOD that are incorrect when applied to a low luminosity sample of galaxies. In the next section, we investigate possible extensions to our vanilla HOD.

\section{Extensions of the HOD}\label{decorated}
\subsection{Spatial bias}
\label{sec:spatial}
In our vanilla HOD model, we assume that each central galaxy lives at the centre of its halo, and that satellite galaxies trace the spatial distribution of dark matter within the halo. However, it is possible that these assumptions are incorrect, i.e. that galaxies exhibit spatial bias. More specifically, central spatial bias occurs when the central galaxy is not located at the centre of its halo, and satellite spatial bias occurs when the satellite galaxies do not trace the distribution of dark matter particles within their halo. 
To test for the presence of spatial bias, one option is to add spatial bias parameters to our HOD model and find a new best-fitting model that includes spatial bias. However, a simpler alternative is to remove the potential effects of spatial bias from the hydrodynamic simulation. If doing this yields better agreement between clustering statistics from our DMO+HOD mocks and the simulation galaxies, this would indicate that there is spatial bias in the hydrodynamic simulation, and therefore spatial bias parameters will need to be included in any future HOD modelling work to account for the possibility that there is spatial bias present in survey data.

We first test for the presence of central spatial bias. We do this by taking the Illustris and EAGLE galaxies identified as centrals and give them the position of their host halo, which is the position of the particle with the minimum gravitational potential energy. We do this without changing any central velocity information or any satellite galaxy information, in order to isolate the effect of central spatial bias. Thus, if there is any central spatial bias present in the original simulation, this procedure would remove it, yielding better agreement with our HOD model. The results of this show no change for either simulation or sample, indicating that any central spatial bias has a negligible impact on clustering statistics.

We next test for the presence of satellite spatial bias. We do this by taking the galaxies identified as satellites in the hydrodynamic simulations and assigning them the positions of random dark matter particles in their host halo (also in the hydrodynamic simulations). We do this without changing any satellite velocity information or any central galaxy information, in order to isolate the effect of satellite spatial bias. We repeat this process 1000 times, in order to generate 1000 different realizations of our simulation with satellite spatial bias removed. We can therefore generate 1000 different $p$-values for each clustering statistic. Table~\ref{tab:all_pval_table} (rows labeled ``Satellite Spatial Bias'') lists the median $p$-values from these 1000 realizations of our simulation with satellite spatial bias removed. We note that it is possible that placing satellite galaxies on dark matter subhaloes rather than particles would alleviate some of the tension that we see between our HOD and the hydrodynamic simulations. However, traditional HOD models do not use subhaloes, in part because the DMO simulations to which they are applied often do not have high enough resolution to resolve small subhaloes. Therefore, we do not explore the option of placing satellite galaxies on dark matter subhaloes in this analysis, but note that it would be worth investigating in future work.

The third point in each panel of Figure~\ref{fig:pvals} shows these median $p$-values that result from both correcting the DMO halo masses and removing satellite spatial bias from the hydrodynamic simulations. Error bars show the range of $p$-values that correspond to the middle 68\% of our 1000 realizations with satellite spatial bias removed. We can see that the $M_r^{-21}$ samples show either slight improvement or no change after removing satellite spatial bias, while the $M_r^{-19}$ samples show significant improvement. In particular, the projected and redshift-space correlation functions are much improved in the $M_r^{-19}$ samples of both EAGLE and Illustris. From these results, we can conclude that the galaxies in EAGLE and Illustris do exhibit satellite spatial bias, the effects of which are more prominent when considering low luminosity galaxies. We can also conclude that the effects shown are definitively the results of spatial bias and not a difference in halo profile due to the presence of baryons; if the clustering differences were due to a difference in halo density profile when baryons are included versus when they are not, then giving the satellite galaxies the positions of random dark matter particles in the halo (in the hydrodynamic simulation) would not have a significant effect on clustering. 

The extent and nature of the satellite spatial bias is similar in the two different simulations. In Figure~\ref{fig:illustris_19_dark_all_stats}, it is clear from looking at both the projected and redshift-space correlation functions that Illustris $M_r^{-19}$ galaxies are less clustered on small scales than the DMO+HOD mock galaxies, or in other words, Illustris galaxies are less concentrated than the dark matter. When satellite spatial bias is removed, the satellite galaxies become more concentrated, and are thus a better fit to the HOD on small scales. The picture looks a bit different in Figure~\ref{fig:eagle_19_dark_all_stats}, where EAGLE $M_r^{-19}$ galaxies are less clustered than DMO+HOD mock galaxies on small scales. However, this amplitude difference in the correlation functions extends to large scales and is thus not caused by satellite spatial bias (it is caused by assembly bias, as we will see later). If we examine the slopes of the correlation functions at small scales in Figure~\ref{fig:eagle_19_dark_all_stats}, we see that EAGLE $M_r^{-19}$ galaxies have a shallower slope than DMO+HOD, which means that they are less concentrated within their haloes \citep{2002ApJ...575..587B}, similar to Illustris $M_r^{-19}$ galaxies.

Despite the improvement that we see in Figure~\ref{fig:pvals} when removing spatial bias, many clustering statistics for the $M_r^{-19}$ samples are still not well predicted by our HOD model, even after correcting the halo mass function and removing satellite spatial bias from the simulations. This is especially true for EAGLE $M_r^{-19}$ galaxies, where all statistics except number density and group multiplicity function still show a significant discrepancy between hydrodynamic and DMO+HOD galaxies.

\subsection{Velocity bias}
\label{sec:velocity}

The vanilla HOD model also assumes that each central galaxy moves with the mean velocity of its halo (i.e. there is no central velocity bias), and that satellite galaxies trace the velocity distribution of dark matter within their halo (i.e. there is no satellite velocity bias). Once again, it is possible that these assumptions are incorrect, due to the effects of phenomena such as mergers, dynamical friction, and tidal stripping.

To test for the presence of central velocity bias, we take the Illustris and EAGLE galaxies identified as centrals and assign them the velocity of their host halo. By doing this, we are removing the possibility that the central galaxy might not be at rest with respect to its host halo. In Illustris, this is the sum of the mass weighted velocities of all particles/cells in the group, multiplied by $1/a$. (In EAGLE, the velocity of the parent halo is not provided, so this test is not possible. Central galaxies already have the velocity of the central subhalo.) As in the case of central spatial bias, removing central velocity bias has a negligible effect on the clustering statistics we consider.

To remove satellite velocity bias, we take the hydrodynamic simulation galaxies identified as satellites and assign them the velocities of random dark matter particles in the halo. We do this in combination with other effects (e.g. central velocity bias, central spatial bias, satellite spatial bias). In other words, we take the central galaxy and give it the position and velocity of its host halo, and we take satellite galaxies and give them the positions and velocities of randomly chosen dark matter particles in the halo, so that all spatial and velocity bias has been removed from the simulation galaxies. We repeat the random selection of dark matter particles 1000 times, so that we ultimately generate 1000 different realizations of the simulation galaxies after removing all spatial and velocity bias. The results of this are shown in Table~\ref{tab:all_pval_table}, where the $p$-values given are the median of 1000.

The fourth point in each panel of Figure~\ref{fig:pvals} shows these median $p$-values that result from correcting DMO halo masses and removing spatial and velocity bias from the hydrodynamic simulations. Once again, error bars show the range of $p$-values that correspond to the middle 68\% of our 1000 realizations with satellite spatial and velocity bias removed. The figure shows that removing velocity bias provides an additional improvement for our clustering statistics for the $M_r^{-19}$ samples. In particular, the Illustris $M_r^{-19}$ sample shows significant improvement in the void probability function and slight improvement in all other clustering statistics. All statistics now show no significant discrepancy between the hydrodynamic galaxies and our DMO+HOD model. The EAGLE $M_r^{-19}$ sample shows improvement in the redshift-space correlation function, as well as both counts-in-cells statistics. It is to be expected that number density does not change when spatial and velocity bias are removed, because the number of galaxies is not affected. Additionally the projected correlation function is by design not affected by velocity, so it is not surprising that there is no change after removing velocity bias. Despite these improvements, the differences between the statistics of EAGLE $M_r^{-19}$ galaxies and the DMO+HOD model are still highly significant. 

At this point, after removing all spatial and velocity bias from our simulations, all statistics are well predicted ($<2\sigma$ tension) by our HOD model for the Illustris $M_r^{-19}$ sample, while the number density and group multiplicity function are well predicted ($<2\sigma$ tension) for the EAGLE $M_r^{-19}$ sample. However, the correlation functions and counts-in-cells statistics are still not well predicted for EAGLE $M_r^{-19}$. This indicates the possibility that the number of galaxies in a halo may depend on a halo property other than mass, such as age or concentration. This will be discussed in the next section. 

\subsection{Assembly/secondary bias}
\label{sec:assembly}
Halo assembly/secondary bias is the phenomenon whereby halo clustering depends on a secondary parameter, such as age or concentration, at fixed halo mass \citep[e.g.,][]{2005MNRAS.363L..66G,2006ApJ...652...71W,2018MNRAS.475.4411S}. If the number of galaxies in a halo depends on this secondary parameter, the clustering of galaxies will inherit this additional halo clustering, a phenomenon known as galaxy assembly bias \citep[e.g.,][]{2007MNRAS.374.1303C,2014MNRAS.443.3044Z}. Galaxy assembly bias could be present in Illustris or EAGLE, but it is explicitly not present in our DMO+HOD model. We now
remove any effects of assembly bias from our hydrodynamic simulation galaxies, with the understanding that if this procedure improves our ability to predict clustering statistics with our DMO+HOD model, this is an indication that future HOD modelling should incorporate parameters that deal with assembly bias.

To remove the presence of assembly bias from our simulation galaxies, we identify pairs of haloes with similar masses, and swap the positions and velocities of their galaxies. This is done after already removing all spatial and velocity bias. In other words, we first generate 1000 realizations of the simulation galaxies after removing spatial and velocity bias (as described above), and then exchange galaxies in haloes of similar mass. When we exchange galaxies in pairs of haloes, we shift the galaxy positions by the difference in halo centre positions, so that a galaxy is in the same position relative to the halo centre, but the halo centre has been switched. For the velocities, we take the peculiar velocity of a galaxy and subtract the mean halo velocity, thus putting the galaxy in the frame of the halo. We then add this velocity to the velocity of the new halo to get the new velocity of the galaxy. In other words, we keep the velocity of the galaxy in the frame of its halo the same, and simply give it a new halo velocity. We use four different combinations of halo pairs, ultimately resulting in 4000 realizations of our simulation galaxies after removing all spatial, velocity, and assembly bias. 

This procedure of exchanging galaxies in haloes of similar mass effectively removes assembly bias from our data because it nullifies any environmental effects on the number of galaxies in each halo. If the number of galaxies in each halo was already only dependent on halo mass, then this procedure should not produce any change in clustering statistics. However, if the number of galaxies in a halo had a dependence on a property other than halo mass, then swapping galaxies in haloes with similar masses would remove the effect of this phenomenon on our clustering statistics. The results of this are detailed in Table~\ref{tab:all_pval_table}. Once again, the $p$-values given are the median of many realizations (in this case 4000).

The last point in each panel of Figure~\ref{fig:pvals} shows these median $p$-values that result from removing assembly bias (in addition to correcting the HMF and removing all spatial and velocity bias). Once again, error bars show the range of $p$-values that correspond to the middle 68\% of our 4000 realizations. Removing assembly bias results in all clustering statistics being well predicted by our HOD for both simulations and luminosity samples. In the $M_r^{-21}$ samples, all clustering statistics were already well predicted, so there is very little change. More importantly, in the $M_r^{-19}$ samples, there are slight improvements in all clustering statistics for the Illustris galaxies, and there are major improvements for the correlation functions and counts-in-cells statistics for the EAGLE galaxies. Of particular note is the void probability function for the EAGLE $M_r^{-19}$ sample, which remained below $5\sigma$ until assembly bias was removed, at which point it reached $1\sigma$ confidence that the HOD model is a good fit to the simulation. This agrees with the results of \citet{2016MNRAS.460.3100C}, who detected galaxy assembly bias in the EAGLE simulation, and found that the signature of assembly bias was stronger for low mass galaxies. This is also consistent with the results of \citet{2006ApJ...647..737T}, which suggested that VPF is sensitive to the presence of assembly bias. More recently, \citet{2019MNRAS.tmp.1686W} also showed that counts-in-cells statistics can be powerful probes of assembly bias. 

\section{Summary and Discussion} \label{summary}

In this work, we have examined the validity of using halo occupation distribution modelling to reproduce galaxy clustering statistics. Halo models provide a simple and computationally inexpensive way to investigate the connection between galaxies and their dark matter haloes, but they rely on the assumption that the role of baryons can be easily parametrized in the modelling procedure. Using two different hydrodynamic simulations, Illustris-2 and EAGLE, we have investigated the accuracy of using a simple five-parameter HOD to reproduce clustering when applied to a high luminosity sample of galaxies as well as a low luminosity sample. The HOD was fit to each simulation and luminosity sample separately, and applied to haloes from the dark matter only counterparts of Illustris and Eagle to create mock galaxy catalogues. Our clustering statistics were measured in the same way on our simulation galaxies as they were on our mock catalogues. Our main results are the following:

\begin{table*}
    \centering
	\caption{Our fits to the halo mass ratios in Illustris-2 and EAGLE, as well as TNG100-2. In the third column, $x$ is equal to $\mathrm{log} M_\mathrm{halo}$.}
	\label{tab:fit_table}
	\begin{tabular}{ccc}
		\hline
		Simulation & Mass Range & $M_\mathrm{halo,Hydro}/M_\mathrm{halo,DMO}$ \\
		\hline
		Illustris-2 & $1.00 \times 10^{11} < M < 9.57 \times 10^{12}$ & $-0.10771x + 2.21907$ \\
		Illustris-2 & $M > 9.57 \times 10^{12}$ & $0.07174x - 0.10774$ \\
		\hline
		EAGLE & $M > 1.00 \times 10^{11}$ & $0.05956x + 0.16413$ \\
		\hline
		TNG100-2 & $1.00 \times 10^{10} < M < 2.74 \times 10^{12}$ & $-0.10171x^2 + 2.37863x - 12.97684$ \\
		TNG100-2 & $2.74 \times 10^{12} < M < 1.06 \times 10^{13}$ & $0.00189x + 0.84450$ \\
		TNG100-2 & $M > 1.06 \times 10^{13}$ & $0.09429x - 0.35479$ \\
		\hline
	\end{tabular}
\end{table*}

\begin{itemize}
    \item Overall, the vanilla HOD model is more successful when applied to a high luminosity sample of galaxies than it is when applied to a low luminosity sample of galaxies.
    \item The simple five-parameter HOD model is able to accurately (within $3\sigma$ tolerance) reproduce correlation functions, the group multiplicity function, the void probability function, and the singular probability function, for the high luminosity sample of galaxies in both Illustris and EAGLE, as well as the number density in EAGLE.
    \item In our $M_r^{-21}$ Illustris and EAGLE samples, the errors among our 1000 mocks are much larger than the cosmic variance errors from the 400 SDSS-like mocks. In other words, an SDSS size $M_r^{-21}$ survey would perhaps be sensitive to clustering differences that we are unable to detect in our analysis. In our $M_r^{-19}$ Illustris and EAGLE samples, the errors among our 1000 mocks are smaller than the cosmic variance errors from the 400 SDSS-like mocks. This means that a survey with the precision of SDSS might not be able to detect the differences that we find between hydrodynamic galaxies and the HOD model. A future survey like the Dark Energy Spectroscopic Instrument \citep[DESI,][]{2016arXiv161100036D}, however, will have better precision than the SDSS due to its larger volume, allowing it to potentially detect these small differences in clustering measurements.  
    \item In general, the halo mass function is shifted to higher masses when baryons are not included, resulting in an over prediction of galaxy number density when an HOD is applied to the haloes from the dark matter only simulations. After correcting the dark matter only halo mass function, the vanilla HOD model is able to accurately reproduce all clustering statistics in the high luminosity sample of galaxies in both Illustris and EAGLE. It also able to accurately reproduce galaxy number density in both low luminosity samples.
    \item Even after correcting the halo mass function, the vanilla HOD model is still unable to accurately (within $3\sigma$ tolerance) reproduce most of the other five clustering statistics for the low luminosity samples of galaxies in Illustris-2 and EAGLE. However, after removing the potential effects of spatial, velocity, and assembly bias from the galaxies in the original simulations, the HOD model (with mass function correction) is able to accurately reproduce all clustering statistics in both samples and both simulations.
\end{itemize}

These results demonstrate the prominent differences between the EAGLE and Illustris simulations, in terms of the ways that baryons affect halo masses and galaxy clustering. For example, the EAGLE and Illustris simulations are very different in terms of the amount of spatial, velocity, and assembly bias they exhibit. Additionally, neither EAGLE nor Illustris reproduces the galaxy luminosity function from the SDSS. Therefore, we cannot use the results from our analysis of the clustering in these two hydrodynamic simulations to draw conclusions about galaxy clustering in the real Universe. Because of this, we do not attempt to infer the true amounts of spatial, velocity, and assembly bias in the real Universe based on this work, but rather recommend that any future work involving HOD modelling should include free parameters for these biases. Moreover, our work suggests that future work aiming to use HOD modelling to study cosmology would benefit from focusing on high luminosity galaxy samples, which seem to be less affected by the aforementioned biases.

Additionally, different clustering statistics are sensitive to different biases. For example, the void probability function seems to be particularly sensitive to the presence of assembly bias, while the redshift space correlation function is sensitive to satellite velocity bias, as can be seen in the low luminosity sample of EAGLE galaxies. Therefore, properly constraining HOD parameters (especially when including spatial, velocity, and assembly bias parameters), necessitates measuring several different clustering statistics.

Of particular note is the difference in how baryons alter the halo mass function between the two different simulations. Any future work hoping to use HOD modelling will have to first correct the dark matter only haloes by shifting the mass function to lower masses, so that it more closely resembles what the mass function would look like with baryons included in the simulation. However, the exact nature of this correction to the halo mass function clearly depends upon which hydrodynamic prescriptions are regarded as the truth. The large difference that we see between the two simulations in Figure~\ref{fig:halo_masses} demonstrates the extent to which mass corrections depend on the details of supernova and AGN feedback physics. This result is somewhat alarming because, unlike the other biases we examine in this study, the effect of baryons on the halo mass function cannot be easily parametrized, making it unclear how one must proceed with halo modelling of observed clustering measurements.

At a minimum, we recommend that future halo modelling efforts repeat their analyses a couple times, applying different corrections to the dark matter only halo masses. This will provide a rough estimate of the systematic uncertainty due to baryonic effects on the halo mass function. For example, if a study finds strong evidence of assembly bias when applying no correction to the halo masses, but then the evidence disappears when the analysis is repeated using a mass correction, one should not claim any detection of assembly bias. To facilitate such a procedure, we fit simple functions to the mass corrections shown in Figure~\ref{fig:halo_masses}. In the case of EAGLE we fit a single line, while for Illustris we fit a broken line. These fits are shown as dashed lines in Figure~\ref{fig:halo_masses}. In Table~\ref{tab:fit_table} we list the parameters for these fits to the mass corrections in Illustris and EAGLE.

We have tested these fits and confirmed that they produce the same results as doing the full abundance matching correction that we performed in our analysis. Additionally, we present fits to the same mass correction in IllustrisTNG \citep{2018MNRAS.477.1206N,2018MNRAS.475..648P,2018MNRAS.475..624N,2018MNRAS.480.5113M,2018MNRAS.475..676S}. TNG is more recent than both Illustris and EAGLE, and makes use of updated feedback mechanisms, which results in a halo mass correction that is different than what we see in either Illustris or EAGLE. We make no assumptions about which of these simulations produces the correct relationship between the masses of their hydrodynamic and DMO haloes, but we recommend that future halo modelling work makes use of one or more of these corrections. 

Rather than viewing these results as evidence that dark matter only simulations are insufficient for halo modelling and should thus not be used to study galaxy clustering, we interpret these results as confirmation that there is no consensus among hydrodynamic simulations. Therefore, dark matter only simulations and halo models are still very relevant tools for investigating the galaxy-halo connection, as long as the halo model is given sufficient freedom, and the effect of baryons on the halo mass function is accounted for. \\ \\ \\

\section*{Acknowledgements}
We acknowledge the Virgo Consortium for making their simulation data available. The EAGLE simulations were performed using the DiRAC-2 facility at Durham, managed by the ICC, and the PRACE facility Curie based in France at TGCC, CEA, Bruy\`eres-le-Ch\^atel.
Some of the computational facilities used in this project were provided by the
Vanderbilt Advanced Computing Center for Research and
Education (ACCRE). This research has made use of NASA's Astrophysics Data System, as well as \texttt{Python} (\texttt{https://www.python.org/}), the \texttt{IPython} package \citep{PER-GRA:2007}, \texttt{SciPy} \citep{jones_scipy_2001}, \texttt{NumPy} \citep{van2011numpy}, and \texttt{matplotlib}, a Python library for publication quality graphics \citep{Hunter:2007}. These acknowledgements were compiled using the Astronomy Acknowledgement Generator (\texttt{http:
//astrofrog.github.io/acknowledgment-generator/}).



\bibliographystyle{mnras}
\bibliography{hod}


\bsp	
\label{lastpage}
\end{document}